\pdfoutput=1
\documentclass[%
aps,
reprint,
superscriptaddress,
]{revtex4-1}

\usepackage{graphicx}
\usepackage{dcolumn}
\usepackage{bm}
\usepackage{epstopdf}
\usepackage{xr}
\usepackage{amsmath}
\usepackage{color}
\usepackage[normalem]{ulem}
\usepackage{wasysym}
\usepackage{siunitx}
\usepackage{physics}
\usepackage{lineno}
\usepackage[T1]{fontenc}
\usepackage{booktabs}
\usepackage[section]{placeins}
\usepackage{textcomp}
\usepackage{amssymb}
\usepackage[english]{babel}

\usepackage{hyperref}
\hypersetup{colorlinks=true,%
citecolor=blue,%
filecolor=black,%
linkcolor=black,%
urlcolor=blue
}

\makeatletter

\newcommand{\Rmnum}[1]{\expandafter\@slowromancap\romannumeral #1@}
\makeatother

\begin{document}

\markboth{\jobname}{\jobname .tex}

\title{Dynamics of an interfacial bubble controls adhesion mechanics in a van der Waals heterostructure}
\author{L.D. Varma Sangani}
\email{sldvarma.999@gmail.com}
\affiliation{Department of Condensed Matter Physics and Materials Science, Tata Institute of Fundamental Research, Homi Bhabha Road, Mumbai 400005, India.}
\author{Supriya Mandal}
\affiliation{Department of Condensed Matter Physics and Materials Science, Tata Institute of Fundamental Research, Homi Bhabha Road, Mumbai 400005, India.}
\author{Sanat Ghosh}
\affiliation{Department of Condensed Matter Physics and Materials Science, Tata Institute of Fundamental Research, Homi Bhabha Road, Mumbai 400005, India.}
\author{Kenji Watanabe}
\affiliation{National Institute for Materials Science, 1-1 Namiki, Tsukuba 305-0044, Japan.}
\author{Takashi Taniguchi}
\affiliation{International Center for Materials Nanoarchitectonics, National Institute for Materials Science, 1-1 Namiki, Tsukuba 305-0044, Japan.}
\author{Mandar M. Deshmukh}
\email{deshmukh@tifr.res.in}
\affiliation{Department of Condensed Matter Physics and Materials Science, Tata Institute of Fundamental Research, Homi Bhabha Road, Mumbai 400005, India.}

\begin{abstract}

  {2D van der Waals heterostructures (vdWH) can result in novel functionality that crucially depends on interfacial structure and disorder. Bubbles at the vdWH interface can modify the interfacial structure. We probe the dynamics of a bubble at the interface of a graphene-hBN vdWH by using it as the drumhead of a NEMS device because nanomechanical devices are exquisite sensors. For drums with different interfacial bubbles, we measure the evolution of the resonant frequency and spatial mode shape as a function of electrostatic pulling. We show that the hysteretic detachment of layers of vdWH is triggered by the growth of large bubbles. The bubble growth takes place due to the concentration of stress resembling the initiation of fracture. The small bubbles at the heterostructure interface do not result in delamination as they are smaller than a critical fracture length. We provide insight into frictional dynamics and interfacial fracture of vdWH.}
\end{abstract}

\maketitle
\vspace{1cm}

Nano-electromechanical systems (NEMS) made using two-dimensional (2D) materials are remarkable due to their low mass, high strength, and large tunability\cite{Lee2008MeasurementGraphene,Yu2021MechanicallyMembranes}. A wide range of NEMS with different classes of 2D materials, including graphene, hexagonal boron nitride (hBN), MoS$_2$, CrI$_3$, have been successfully demonstrated \cite{Bunch2007ElectromechanicalSheets, Mathew2016DynamicalDrums,Zheng2017HexagonalMotion, Lee2018ElectricallyRange,Jiang2020ExchangeAntiferromagnets}. Their versatile properties are promising for gas and chemical sensors \cite{Koenig2012SelectiveGraphene}, pressure sensors\cite{Patel2016LowSensing}, magnetic sensors\cite{Jiang2020ExchangeAntiferromagnets}, mass sensors \cite{Chaste2012AResolution}, and switches \cite{Liu2014LargeSwitches}. In addition, different 2D materials can be stacked to make van der Waals heterostructures (vdWH) that exhibit emergent properties\cite{Geim2013VanHeterostructures}.

Graphene is a widely studied 2D material with its high mechanical strength and interesting electronic properties \cite{Lee2008MeasurementGraphene}. hBN with a large bandgap of 5.9~eV \cite{Watanabe2004Direct-bandgapCrystal} is an excellent substrate and provides encapsulation for graphene and other 2D materials due to its flatness, thermal stability, and absence of dangling bonds \cite{Dean2010BoronElectronics}. The vdW graphene-hBN heterostructure with superior electronic properties has led to renewed studies of novel quantum Hall effects. The differing lattice constants of graphene and hBN result in a moiré superlattice\cite{Yankowitz2019VanNitride}. This aspect is utilized in twistronic devices that show dramatic tuning of electronic properties due to the strong interlayer interaction and low disorder over the superlattice length scale \cite{Cao2018CorrelatedSuperlattices}. However, local strain can cause variation in the twist angle, which can adversely affect the device performance \cite{Uri2020MappingGraphene}. Hence, a detailed study of local strain and its dependence on the interface conditions in a vdWH is essential.

Mechanical properties of 2D material interfaces have been studied using NEMS devices \cite{Ye2021UltrawideResonators,Kim2020StochasticInterfaces,Kim2018Nano-electromechanicalBimorphs,Kumar2020CircularHeterostructures}, STEM imaging \cite{Yu2021DesigningHeterostructures,han2020ultrasoft}, AFM-based techniques\cite{Frank2007MechanicalSheets}, blister tests\cite{Cao2014AGraphene,Wang2019BendingMaterials}, use of stochastic and defect nucleated bubbles \cite{Zong2010DirectNanoparticles,Yoon2012DirectProcess}. These experiments provide important information about the interface and its adhesion properties at the nanoscale; however, they are quasi-static in nature. Combining NEMS and vdWH allows probing dynamic properties of interfacial interaction. There is an increasing interest in vdWH based NEMS to understand the mechanical properties of heterostructures, the nature of their interfaces \cite{dai2020mechanicsopportunities}, and possible interdependence between their mechanical and electronic properties \cite{Pereira2009StrainStructure}. Recent reports on interface interactions of graphene-MoS$_2$ \cite{Kim2018Nano-electromechanicalBimorphs}, CVD-grown bilayer graphene \cite{Kim2020StochasticInterfaces,Ferrari2021DissipationResonators}, graphene-hBN \cite{Kumar2020CircularHeterostructures}, and graphene encapsulated NbSe$_2$ \cite{Will2017HighResonator} heterostructures utilizing NEMS resonator show the capability of NEMS to investigate the interface properties. 

In this work, we study graphene-hBN heterostructure based NEMS to probe their electromechanical response and its dependence on the interface quality. A systematic study of the role of bubbles in such an interface provides several useful information related to the interfacial dynamics and has some applications. For example, bubbles have been predicted to produce gauge fields (pseudo magnetic fields) that can have interesting consequences for quantum transport \cite{jia2019programmable,de2011aharonov}. Additionally, understanding the effects of bubbles on interfacial stress can provide guidance to what sort of bubble dimension can affect other measurements, like transport, on heterostructure devices. We prepare graphene-hBN van der Waals heterostructures with a different configuration of bubbles and use them as drumheads for circular resonators. We focus on three classes of drum variants in this work -- a clean graphene-hBN interface with no bubbles, few small ($\approx$~100~nm size) bubbles, and a single large ($\approx$~700~nm) bubble at the interface. We observe noticeable differences in the frequency dispersion and the mode shapes of these devices with gate voltage. We understand the device response using complementary finite element modelling (FEM). Our results show that the frequency dispersion with an electrostatic gate and resonant excitation is due to the dynamic growth of the bubble and the eventual detachment of graphene from hBN is due to fracture. Griffith's critical length \cite{griffith_vi_1921} physics is borne out as the smaller bubbles (~$\approx 100$~nm) do not result in delamination while the single larger bubble (~$\approx~700$~nm) causes fracture induced delamination due to critical accumulation of stress; these experimental observations are consistent with our calculations. Our experiment probes the dynamics of this vdWH interface for the first time and we also observe the hysteretic nature of the bubble growth dynamics at nanoscale.

\begin{figure*}
\centering
\includegraphics[scale=1]{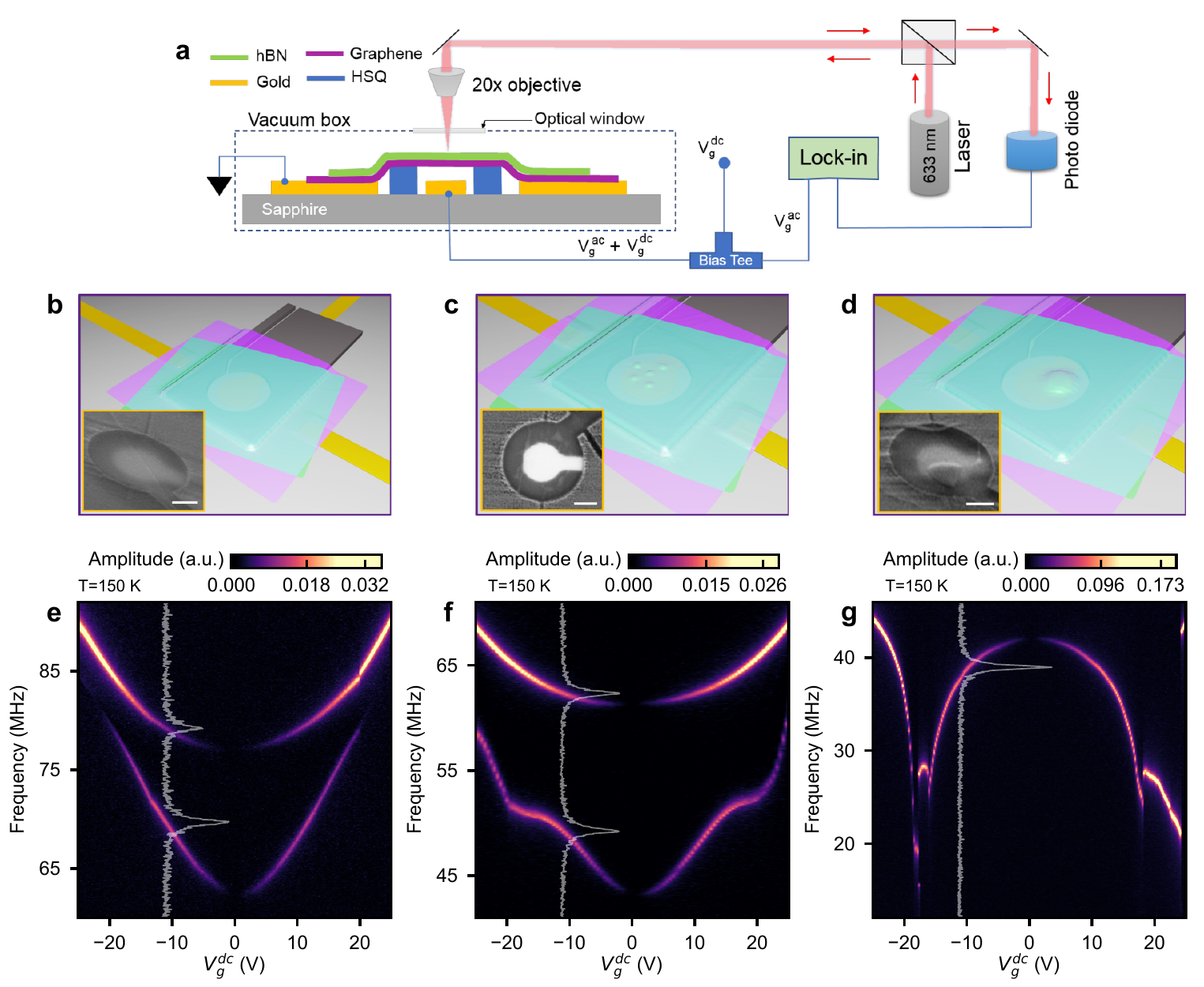}
\caption{\textbf{Schematic of vdWH NEMS and their  frequency dispersion.} (a) Schematic diagrams of the drum resonator cross-section under electrical actuation and optical detection measurement scheme, the laser wavelength is 633~nm. (b, c, d) Schematics of G-hBN drums with no bubble, with few small bubbles and with a large bubble respectively. Inset images show SEM images of corresponding drums. The pink color flake is few-layer graphene and the green color flake is few-layer hBN and the scale bar in the inset SEM images is equal to 1 \textmu m. Both graphene and hBN thicknesses are around 10~nm, details of fabrication provided in section S1 of the supplementary information. (e, f, g) 2D color plots of frequency dispersion of the drums shown in b, c, \&d (D1, D2 $\And$D3) respectively, measurements were carried out at $T=150$ K. The color plots in Figure \ref{fig:Schematics & 2D plots}e-g represent variation in the reflected laser signal, which is proportional to the amplitude of the membrane motion. The white line plot in the color plots represents the frequency scan at $V_g^{\textrm{dc}}=-11$~V for the corresponding 2D plots. More line slices at different gate voltages are shown in supplementary information (Figure S6).}
\label{fig:Schematics & 2D plots}
\end{figure*}

Figure \ref{fig:Schematics & 2D plots}a shows the schematic of the cross-sectional view of the graphene-hBN (G-hBN) drum resonator. The sequence of layers, few-layer hBN on top and few-layer graphene at the bottom, as shown in Figure \ref{fig:Schematics & 2D plots}a, is critical for our study as the electrostatic actuating force acts on graphene because it has mobile charges while hBN is a dielectric with no free charges. This imparts a pulling action across the hBN and graphene interface. The schematic of drums of three different kinds D1, D2, and D3 are shown in Figure \ref{fig:Schematics & 2D plots}b-d, respectively. Inset images of Figure \ref{fig:Schematics & 2D plots}b-d show the SEM images of the respective drums. D1 has a clean interface (no visible bubbles under SEM), D2 has few bubbles (each having an average diameter $\approx$~100~nm), and D3 has a big bubble (diameter $\approx$~700~nm) at the G-hBN interface. The formation of the bubbles at a 2D heterostructure interface in the 2D material stacking process is well known \cite{Khestanova2016UniversalHeterostructures}. It is stochastic and strongly depends on environmental conditions and the stacking process. The origin of these bubbles is either a difference in local strains, a residue, or air trapped at the interface. We make many devices and choose the device in these three subcategories for detailed studies presented here.

The nanomechanical resonant excitation measurement scheme is shown in Figure \ref{fig:Schematics & 2D plots}a. We actuate the drum with electrostatic force by applying a combined DC and RF voltage to the gate \cite{Bunch2007ElectromechanicalSheets}. We monitor the mechanical resonance of the drums by measuring the reflected laser interference signal from the drum cavity.

Figure \ref{fig:Schematics & 2D plots}e-g show the dispersion of resonance frequency of the hetero-drums D1, D2, and D3 with varying gate voltage at temperature $T=150$~K. We choose $150$~K primarily to avoid the phase transition of gases (mostly N$_2$ and O$_2$) inside the bubble which may complicate the dynamics. Data at room temperature and additional data of D1 and D2 are provided in sections S2 and S3 of the supplementary information. The gate voltage ($V_g^{\textrm{dc}})$ sweep direction in these plots is from $V_g^{\textrm{dc}}= -25$~V to 25~V; we refer to this as the forward direction. All the drums showed significant positive resonant frequency tuning with gate voltage.  Such large tuning of resonance frequency has been recently reported in clean MoS$_2$/graphene heterostructure as well\cite{Ye2021UltrawideResonators}. We see two strong modes in drums D1 and D2, and only one in drum D3 in the measurement window. Though both the modes in D1 and the upper mode in D2 show usual gate dispersion \cite{Mathew2016DynamicalDrums}, some unusual features show up in the lowest mode of D2 and D3. These features can be ascribed to the presence of the bubbles in the G-hBN interface, and we focus on them next. 

To perform a comparative study of the interfacial bubbles, we take the D1 response as a reference, and compare that of D2 and D3 with D1. Figure \ref{fig:FWHM, Mode maps}a shows the fundamental mode frequency dispersion of the resonators D1 (green curve) and D2 (pink curve) at 150~K. The device D1 exhibits a usual high positive tunability. While the tunability of D2 is similar to D1, it drops in an intermediate voltage range (shown by gray shaded regions). Such a tunability variation in dispersion can occur due to capacitive  softening\cite{Singh2010ProbingResonators}, mode coupling \cite{Mathew2016DynamicalDrums}, tensioning, or heating effect from the laser \cite{VanDerZande2010Large-scaleResonators,Chen2009PerformanceReadout,Guttinger2017Energy-dependentResonators}. The laser heating effect can be ruled out since an electrical detection technique also shows identical frequency dispersion (Figure S8 in the supplementary information). The tensioning and capacitive softening, in general, results in negative or `w’ shaped dispersion which is different from the D2 response. The variations are also very different from an avoided crossing in which a clear gap-opening appears near a crossing of two modes. On the other hand, similar kind of kinks are observed in 2D heterostructure resonators Graphene/MoS$_2$ by Kim et al. \cite{Kim2018Nano-electromechanicalBimorphs} and hBN/Graphene by Kumar et al.\cite{Kumar2020CircularHeterostructures} and these kink features were attributed to interlayer slip; we note that the system of Kim et al.\cite{Kim2018Nano-electromechanicalBimorphs} is different from our study.  The presence of these reduced tunability regions only in the device with a bubble-interface (D2), but not in the clean-interface device (D1), supports the argument of an interlayer slip. The deviation in D2 dispersion at higher gate voltages is the result of a change in strain in the heterostructure due to interlayer slip\cite{Kim2018Nano-electromechanicalBimorphs}. Furthermore, the same tunability (Figure \ref{fig:FWHM, Mode maps}a) and no hysteresis at lower gate voltages (Figure S7) suggest that the interface has negligible influence on D2 response before the slip.

The frequency dispersion of D3 is significantly different from D1 and D2. As shown in Figure \ref{fig:FWHM, Mode maps}b, three distinct regions can be identified with an increasing magnitude of $V_g^{\textrm{dc}}$, the details of which also depend on the sweep direction. Frequency scan slices in these regions (shown in Figure S6) show that the resonator response is linear in all three regions. We first discuss the dispersion when $V_g^{\textrm{dc}}$ is varied from $-25$~V to 25 V. The first region, R3$^\prime$, is characterized by a large positive dispersion ($\approx$ 10 MHz/V) with $V_g^{\textrm{dc}}$ and the frequency decreases rapidly till $V_g^{\textrm{dc}}$ $\approx$ $-17.5$ V. This is followed by an abrupt increase in frequency and the dispersion becomes negative; this marks the start of the region R2$^\prime$. As $V_g^{\textrm{dc}}$ is swept further towards zero, a kink is observed at $-15$ V. This denotes the start of region R1$^\prime$ where the rate of negative frequency dispersion decreases as $V_g^{\textrm{dc}}$ approaches 0 V, reaching a minimum. Although the dispersion in regions R1, R2 and R3 which belong to positive gate polarity are qualitatively similar to their negative polarity counterparts, few important differences can be noticed. Firstly, the transition from R2 to R3 happens with an abrupt jump in frequency, accompanied by a change from negative to positive dispersion. Secondly, the regions R3 and R3$^\prime$ together with regions R2 and R2$^\prime$ are very different in `width’ while the regions R1 and R1$^\prime$ are of the same width. Moreover, the negative dispersion in both the regions R1 and R1$^\prime$ is unusually large with a frequency reduction of 18 MHz; we discuss this aspect in detail later. Note that the device response is asymmetric with respect to polarity of the gate bias contrary to an expected symmetric response, as the electrostatic force ($\propto (V_g^{\textrm{dc}})^2$) is an even function of gate voltage. On repeated bidirectional gate voltage sweep, a strong hysteresis at larger $V_g^{\textrm{dc}}$ values is observed, as shown in Figure \ref{fig:FWHM, Mode maps}c. The statistics of the jumps are presented in Figure \ref{fig:FWHM, Mode maps}d,e. A small spread in the gate voltages corresponding to the frequency jumps can be noted, they show the stochastic nature of the delamination process.

To better understand the drum’s response and the role of the interfacial bubbles, we next examine the variation of dissipation in the system. Figure \ref{fig:FWHM, Mode maps}f-h show the quality factor (Q) of the devices D1, D2, and D3. The Q of D3 (Figure \ref{fig:FWHM, Mode maps}h) shows a deviation at the crossing between regions defined in Figure \ref{fig:FWHM, Mode maps}b. Specifically, there is a gradual decrease in Q as it approaches the boundary of any two regions. Whereas no such trends were observed in Q of the drums D1 and D2 (Figure \ref{fig:FWHM, Mode maps}f,g). The Q gives a measure of energy dissipation rate, the decrease in Q around the transition between two regions can be attributed to an increase in energy dissipation rate. A rapid change in the Q indicates a new channel of dissipation becoming available in the system.

We now map the mode shape in these three regions of device D3 and compare it with D1 and D2. Figure \ref{fig:FWHM, Mode maps}i-n show the spatial mapping of vibrational modes using laser interference technique at different gate voltages and corresponding resonance frequencies (details about the detection technique are provided in section S1 of the supplementary information). The red dashed lines show the overlaid boundary of the drumheads and the blue dashed lines in Figure \ref{fig:FWHM, Mode maps}l-n show the bubble boundary for the drum D3. From the mode maps of D3 (Figure \ref{fig:FWHM, Mode maps}l-n), we notice a considerable change in amplitude in the vicinity of the bubble at resonance corresponding to regions R2 and R3 than in R1. However, we do not see such changes in the mode maps of D1 and D2.

The behavior of the device D3 is complex and many of the features are observed for the first time in 2D resonators (additional device data is presented in section S8 of the supplementary information). We discuss the key observations that point to a role of the interface in the device with a large bubble (D3). Firstly, we observe an unusual frequency dispersion with gate voltage, with the regions R1/R1$^\prime$ and R2/R2$^\prime$ showing a negative dispersion in the resonant frequency. Capacitive softening in nanomechanical systems \cite{Singh2010ProbingResonators,Kozinsky2006TuningResonators}, with high internal tensile stress can result in reduced effective stiffness of the system with $\Delta k \propto -\frac{\partial^2C}{\partial z^2}$ where $k$ is spring constant, $C$ is capacitance and $z$ the spatial coordinate of vibration. However, this cannot explain the large variation ($\approx$ $-4$~MHz/V) observed in our device as negative dispersion is typically small \cite{VanDerZande2010Large-scaleResonators,Chen2009PerformanceReadout}. Additionally, the huge jumps in frequency cannot be explained either by capacitive softening, tensioning, or mode coupling. They are also different from the variations in dispersion observed for D2 (shown in Figure \ref{fig:FWHM, Mode maps}a). Secondly, from the mode maps and Q variation, we find considerable changes in energy and membrane motion when the device resonance moves from one regime to another for the drum with a single large bubble D3. More energy dissipation at the region to region transition and increase in amplitude in the vicinity of the bubble at high gate voltages suggests a picture of detachment of graphene from hBN. We also measure additional devices and find similar qualitative features, shown in section S8 of the supplementary information. Taken together, we see evidence for interfacial dynamics, in particular, a change in the mode shape in the vicinity of the bubble; this suggests detachment. We note that the dynamics we probe is a response to a resonant actuation that is dynamic and fundamentally different from the static studies of bubbles and interfaces done in the past. To gain more insights into the interfacial dynamics of the bubble, we performed detailed FEM simulations for our device geometry which is discussed below.

Figure \ref{fig:COMSOL D3} focuses on the FEM of the device with a single bubble. Figure \ref{fig:COMSOL D3}a shows the schematic of the cross-section image of G-hBN heterostructure across the bubble and Figure \ref{fig:COMSOL D3}b shows the geometry used for FEM. The interlayer interaction is parametrized as a spring constant per unit area, $k_{\textrm{vdW}}^A$, between graphene and hBN membranes; this is depicted in an enlarged image of the interface in Figure \ref{fig:COMSOL D3}a.  Incorporating a realistic finite element model that uses Lennard-Jones potential while also probing vibrational modes was computationally very demanding. We used a harmonic potential, which is a valid approximation in the stable configuration, i.e. when the layers are attached together by van der Waals force, and then study bubble growth parametrically. Using the implications from our experimental results of no separation between hBN and graphene layers in D1 mode mapping up to $V_g^{\textrm{dc}}=25$~V and some slipping in D2 dispersion, we optimize the distributed spring constant using FEM simulation (details are provided in section S9 of the supplementary information) to be $\approx 8 \times 10^{13}$~N/m$^3$. 

Using values of $k_{\textrm{vdW}}^A$ inferred from experiments and simulations, we can calculate the frequency dispersion of the fundamental mode of D3 geometry without bubble (Figure \ref{fig:COMSOL D3}c) as a function of gate voltage using the distributed spring constant and this is shown in Figure \ref{fig:COMSOL D3}e. The simulations show a `w' shaped frequency response which is well known as the result of spring softening. The drum with a single bubble geometry is depicted in Figure \ref{fig:COMSOL D3}d. The bubble region has no spring constant and it is an asymmetric ellipse with a major axis of 600~nm. Figure \ref{fig:COMSOL D3}e shows the simulation result for frequency dispersion as a function of gate voltage. The result of this model is not much different from without bubble and does not mimic all the features of the actual D3 response and needs additional inputs, namely the dynamical response of the drum. As we pointed out earlier, the DC gate voltage tugs on graphene, and AC gate voltage provides dynamical resonant force at the interface.

Now we discuss how the bubble dynamics can be initiated during the resonant forcing of the bubble trapped between the interface of hBN and graphene. Using ideas of fracture dynamics, we discuss how the stress at the edges of the bubble can be very large to initiate bubble growth. Figure \ref{fig:Fracture like picture}a shows the schematic of the D3 drum head, a cross-section view across the red line is shown in Figure \ref{fig:Fracture like picture}b. Figure \ref{fig:Fracture like picture}b is analogous to an elliptical crack in a plate in fracture mechanics and the stress field bends around the bubble \cite{TedL.Anderson2005FractureApplications}; dashed lines represent the stress field due to applied external stress $\sigma$. Maximum stress will be at the bubble edge A ($\sigma_A$) and it is given by the equation, $\sigma_A=2\sigma\sqrt{{R_{bub}}/{\rho}}$, 
where $R_{bub}$ is the bubble radius and $\rho$ is the bubble edge curvature at A\cite{TedL.Anderson2005FractureApplications}. For the device D3, $R_{bub}\approx$ 300~nm and $\rho$ is few nm. From the equation, the stress at the bubble edge A can be much larger compared to applied stress ($\approx$ 16$\sigma$ for $\rho=$50~A$^\circ$). So, bubble growth is possible even at a low voltage applied to the gate electrode as the bubble edges exceed critical stress \cite{griffith_vi_1921}. Now, using Griffith’s simple picture, $\sigma_F= \sqrt{(G_cY)/(\pi a)}$; here $\sigma_F$ is the fracture stress and $G_c$ is the energy release rate, analogous to adhesion energy,  $Y$ is the Young’s modulus and $2a$ is the critical fracture length. From our FEM calculations (section S9 in the supplementary information), we find that the stress around the bubble can be $\sigma_F\approx~10^8$ N/m$^2$; combining the information from simulations and using the known values of adhesion energy ($G_c\approx$~0.2~J/m$^2$) \cite{dai2020mechanicsopportunities}, and $Y$ \cite{Wang2019BendingMaterials},  we can plot and find that the critical bubble dimension is $\approx$~250 nm (critical bubble diameter, $2a \approx 500$ nm) when sufficient stress accumulates at the boundary for the fracture to take place. Detailed calculation are given in the section S10 of the supplementary information.

Next, we examine the FEM simulations by varying $R_{bub}$ as a function $V_g^{\textrm{dc}}$ parametrically. Figure \ref{fig:Fracture like picture}c shows the FEM simulations with bubble growth governed by different bubble growth rates. The bubble growth rate ($m$) denotes the increase in bubble radius per unit change in gate voltage (\textmu m/V). We have assumed a linear bubble growth rate which is a simplifying assumption to compare with experimental data. The actual growth rate can be non-linear. But even for non-linear cases, our simulation can provide useful insight by considering local tangents in growth rate vs gate voltage curve. We notice bubble growth rate significantly modifies the shape of the frequency dispersion. As $m$ increases, positive dispersion reduces and results in large negative dispersion. Results with growth rate, $m=$ 0.06~\textmu m/V resemble the R1 region in the device D3 response. Additionally, the 2D heterostructure interface contains many other features like residue, wrinkles apart from simple bubbles\cite{Jain2018MinimizingPDMS,Sanchez2018MechanicsCrystals}. They may restrict the detachment unlike the bubbles and may result in a reduction in bubble growth rate and also may result in sudden detachment or slip of graphene from hBN. Figure \ref{fig:Fracture like picture}d shows the FEM simulation with a growth rate $m=0.06$~\textmu m/V including slips at different voltages, labelled in the plot with numbers 2 and 4 (details of this simulation provided in section S9 of the supplementary information). From Figure \ref{fig:Fracture like picture}d, a slip of 5~nm at 16.3~V (label 2) and 50~nm at 22.3~V (label 4) resulted in a 3.5~MHz and 20.8~MHz jump in frequency respectively. Figure \ref{fig:Fracture like picture}e represents the slice color plot of the eigenmode of the drum in the FEM simulation. Reduced tunability in region R2 is the result of slow bubble growth before the large jump. After complete detachment of graphene, positive dispersion in region R3 can be understood to be due to reduced tension in the graphene membrane. The hysteretic behaviour observed in the experimental data is shown in Figure \ref{fig:FWHM, Mode maps}c and is analogous to the hysteresis in the AFM approach and retraction curves. In our system, this is due to the detachment and attachment due to two membranes. Next, we provide further evidence behind the detachment and attachment of graphene and hBN.

In Figure \ref{fig:Fracture like picture}f, we show the response for the device with a single bubble over a large frequency range compared to Figure \ref{fig:Schematics & 2D plots}g. We see similar jumps in the frequency response. However, the  additional key feature is the emergence of new modes, around $\approx$~75~MHz, in region R3$^\prime$ where the hBN and graphene have detached. The emergence of the new mode is consistent with the idea that now the two layers vibrate independently as complete delamination has occurred. 

 We use 2D vdW heterostructure and study the dynamics of the interfacial bubbles using the NEMS devices. We studied three different kinds of devices with no bubbles, few bubbles, and one large bubble. We show bubbles diminish the quality of the interface and beyond a critical size of bubble lead to interfacial fracture. The hysteretic detachment of layers of vdWH takes place in bubble of size $\approx$~700~nm and is triggered by the bubble growth -- the mechanism resembles fracture process. Our observations agree with Griffith's criterion for fracture. Calculations suggest a critical bubble diameter $\approx$~500~nm consistent with our experiment. The dynamics of friction and associated fracture processes are still being understood \cite{gvirtzman_nucleation_2021,malthe-sorenssen_onset_2021}. Our technique of embedding the fracture interface in a NEMS device can provide insights into nanoscale interfaces, along with adhesion and fracture dynamics. Our approach may help in understanding how the fracture front moves as the resonant force amplitude changes. While a lot is known about the quasi-static nature of the fracture, a detailed dynamics study can help in improving the understanding of friction and its nucleation. Combining nanomechanics with in-situ SEM imaging in a stroboscopic manner would also be beneficial for imaging different stages of delamination.

\section{Data avilability}
The COMSOL codes are publicly accessible via Zenodo with the identifier \href{https://doi.org/10.5281/zenodo.5929530}{doi:10.5281/zenodo.5929530}

\section{Acknowledgement} 

We thank Ajay Sood, Vibhor Singh, Rajesh Ganapathy, Manish M. Joglekar, Shankar Ghosh, Prita Pant, and Nagamani Jaya Balila for helpful discussions and comments. We also thank Bhagyashree Chalke, Rudheer Bapat, and Jayesh B Parmar for SEM imaging. We acknowledge the Nanomission grant SR/NM/NS-45/2016 and DST SUPRA SPR/2019/001247 grant along with the Department of Atomic Energy of Government of India 12-R\&D-TFR-5.10-0100 for support. Preparation of hBN single crystals is supported by the Elemental Strategy Initiative conducted by the MEXT, Japan (Grant Number JPMXP0112101001) and JSPS KAKENHI (Grant Numbers 19H05790 and JP20H00354).

\section{{Authors' Contributions}}
L.D.V.S. fabricated the devices. L.D.V.S. and S.G. did the measurements. L.D.V.S., S.M., and M.M.D. analyzed the data and did the finite element simulations. K.W. and T.T. grew the hBN crystals. L.D.V.S. and M.M.D. wrote the manuscript with inputs from everyone. M.M.D. supervised the project


%

\begin{figure*}
\centering
\includegraphics[scale=1]{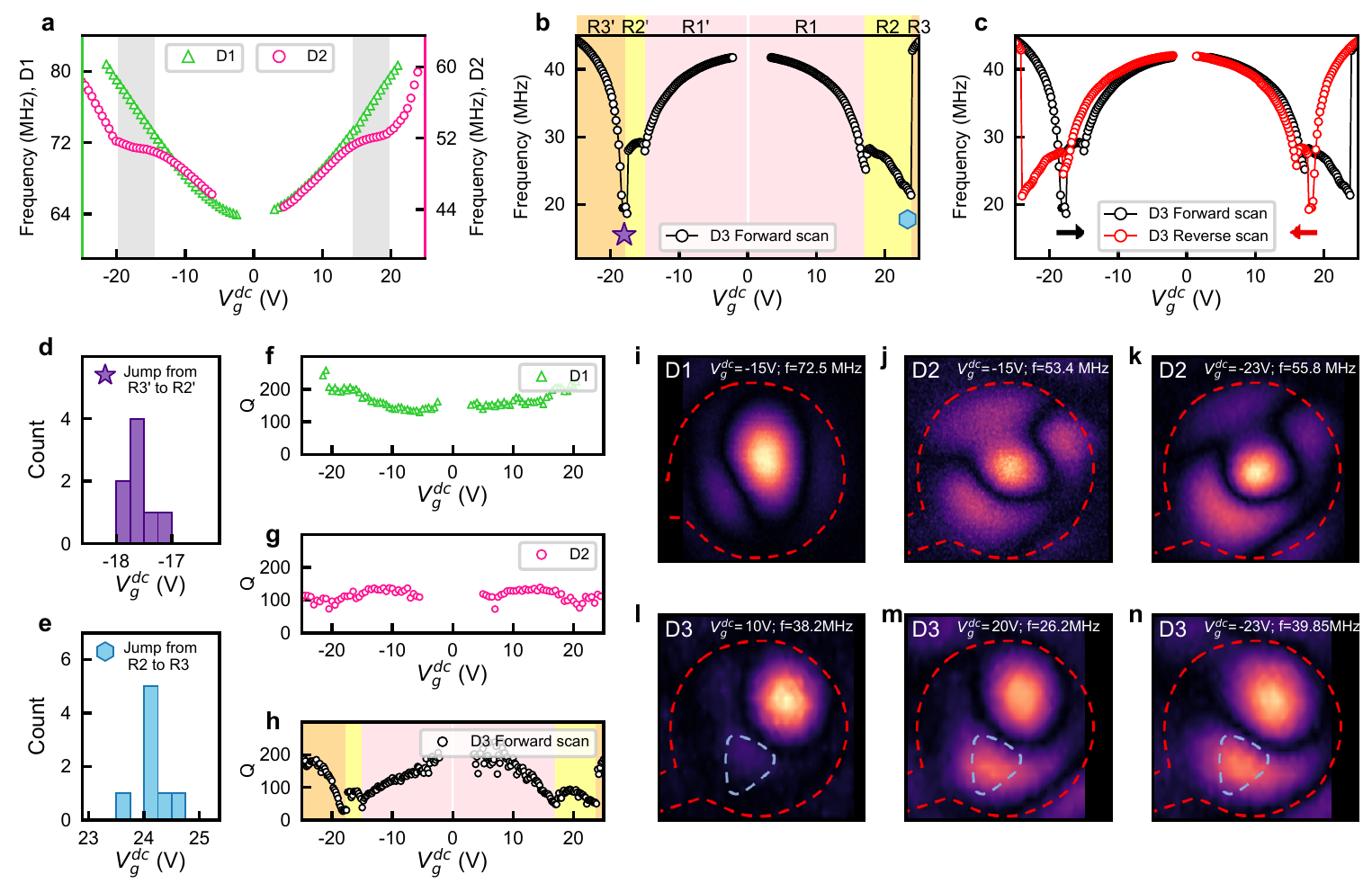}
\caption{\textbf{Resonance frequency dispersion of the devices and spatial mode mapping.} (a) Dispersion of the devices D1 (green) and D2 (pink), kink regions are shaded with gray color. (b) D3 frequency dispersion indicated with a different regions R1, R2, and R3. R1$^\prime$, R2$^\prime$, and R3$^\prime$ are the regions in the negative gate voltage. (c) Hysteresis of the device D3, forward
sweep (black) is from $-25$ to 25~V and reverse sweep (red) is from 25 to $-25$~V, the red and black arrows in (c) indicate the direction of the sweeps. (d,e) Histograms of frequency jump voltages denoted in (b) (with star and hexagon) for 8 cycles. (d) is corresponding to the jump from R3$^\prime$ to R2$^\prime$ (star) and (e) is corresponding to the jump from R2 to R3 (hexagon). (f,g,h) Quality factor (from Lorentzian fit) of the frequency scans of D1, D2, and D3 respectively, shown in (a) and (b). (i-n) Spatial mapping of eigenmodes of D1, D2, and D3 at different voltages using laser interference technique, color plot represents the reflected laser signal proportional to the amplitude of the membrane motion. (i) D1 eigenmodes at $V_g^{\textrm{dc}}=-$15~V (j,k) D2 eigenmodes at $V_g^{\textrm{dc}}=-$15~V and $-$23~V respectively. (l,m,n) D3 eigenmodes at $V_g^{\textrm{dc}}=$10~V, $-$20~V and $-$23~V respectively. Red dashed line in (i-n) represents the drum boundary (diameter 3.5~\textmu m) and blue dashed-line in (l-n) is the boundary of the bubble in D3.}
\label{fig:FWHM, Mode maps}
\end{figure*}

\begin{figure*}
\centering
\includegraphics[scale=1]{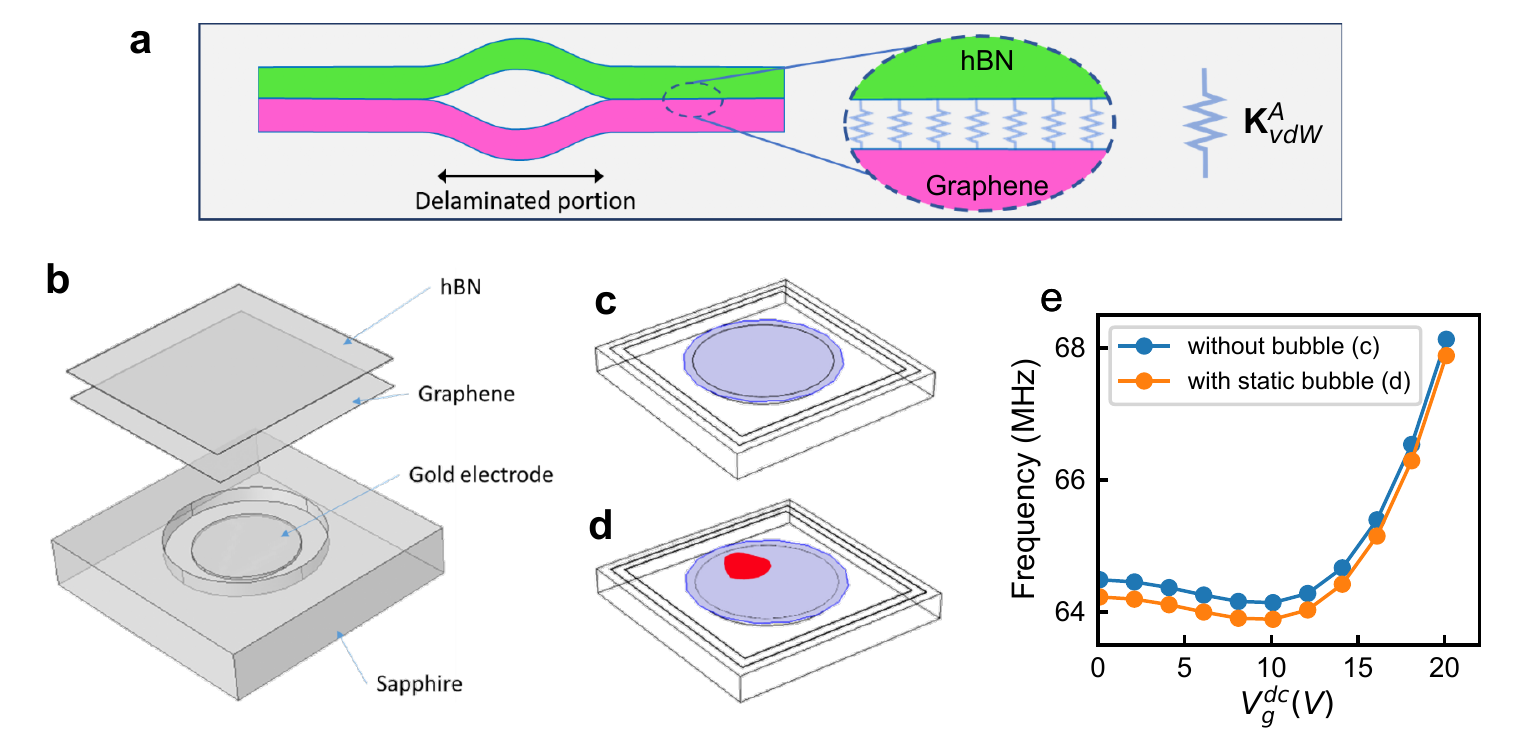}
\caption{\textbf{Bubble at the interface modelled using FEM}. (a) Schematic of the cross-section image of G-hBN heterostructure across the bubble with effective spring system due to vdWs at the interface. $k^A_{\textrm{vdW}}$ is an effective spring constant per unit area due to vdW forces. (b) Drum resonator geometry was used in our FEM simulations. (c,d) D3 FEM-geometries without and with bubble respectively. Blue and red regions are thin elastic layer regions at graphene-hBN interface with $k_{\textrm{vdW}}^A=$ $8\times 10^{13}$N/$m^3$ and 0 N/$m^3$ respectively. (e) FEM simulation results of D3 using geometry (c) and (d).}
\label{fig:COMSOL D3}
\end{figure*}

\begin{figure*}
\centering
\includegraphics[scale=1]{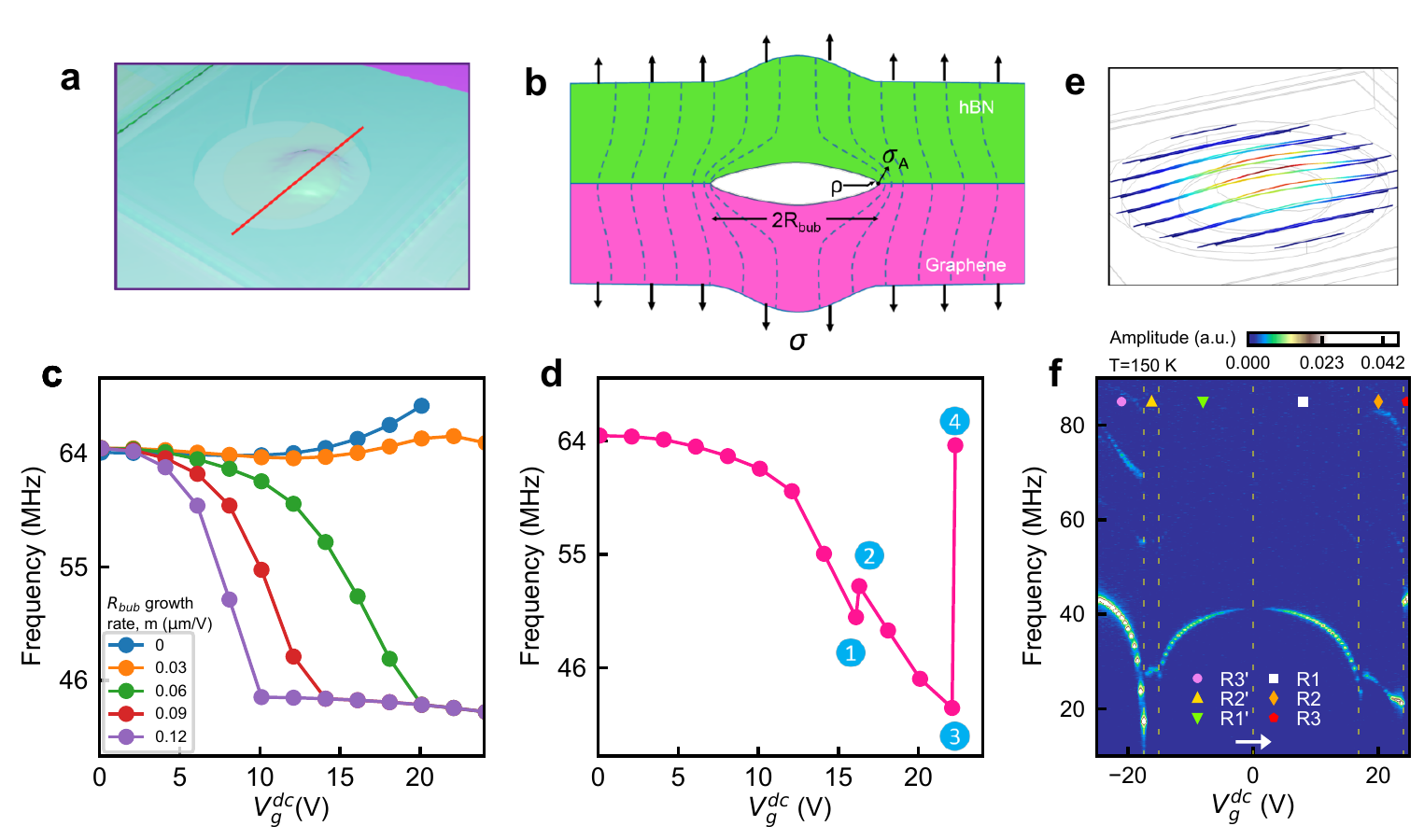}
\caption{\textbf{Fracture at the bubble interface and the emergence of new modes. }(a) Closed up image of D3 schematic shown in Figure \ref{fig:Schematics & 2D plots}d. (b) Schematic of the stress field lines across graphene (pink), hBN (green) heterostructure along the red line indicated in (a). The empty shape at the interface indicates the bubble with radius, $R_{\text{bub}}$. $\rho$, $\sigma$ and $\sigma_A$ are bubble curvature at the interface, applied stress on the heterostructure and stress at the bubble edge respectively. (c) Bubble growth assisted FEM simulation of D3 at different growth rate (increase in bubble radius per unit change in gate voltage), growth is governed equation is, $R_{\textrm{bub}}(V_g^{\textrm{dc}})=R_i+mV_g^{\textrm{dc}}$ for $R_{\textrm{bub}}\leq R_f$. (d) Simulation result at growth rate 0.06~\textmu m/V including 5~nm slip at $V_g^{\textrm{dc}}=16.3$~V and 50~nm slip at $V_g^{\textrm{dc}}=$ 22.3~V. (e) The slice color plot of an eigen mode of the drum in FEM simulation. (f) Frequency dispersion of D3 over frequency range from 10~MHz to 100~MHz, new mode emerged at around 70~MHz in the region R3$^\prime$. More details of this mode are discussed in section S4 in the supplementary information. }
\label{fig:Fracture like picture}
\end{figure*}

\clearpage


\widetext
\begin{center}
\textbf{\large Supplementary Materials: Dynamics of an interfacial bubble controls adhesion mechanics in a van der Waals heterostructure}
\end{center}

\setcounter{section}{0}
\setcounter{equation}{0}
\setcounter{figure}{0}
\setcounter{table}{0}
\setcounter{page}{1}

\renewcommand{\thesection}{S\arabic{section}}
\renewcommand{\theequation}{S\arabic{equation}}
\renewcommand{\thefigure}{S\arabic{figure}}
\renewcommand{\thepage}{S\arabic{page}}
\renewcommand{\bibnumfmt}[1]{[S#1]}
\renewcommand{\citenumfont}[1]{S#1}

 \makeatletter
\def\@fnsymbol#1{\ensuremath{\ifcase#1\or \dagger\or *\or \ddagger\or
   \mathsection\or \mathparagraph\or \|\or **\or \dagger\dagger
   \or \ddagger\ddagger \else\@ctrerr\fi}}
    \makeatother

\section{Fabrication and measurement details}

\subsection{PDMS assisted dry transfer}

\begin{figure}[h]
\centering
\includegraphics[scale=0.15]{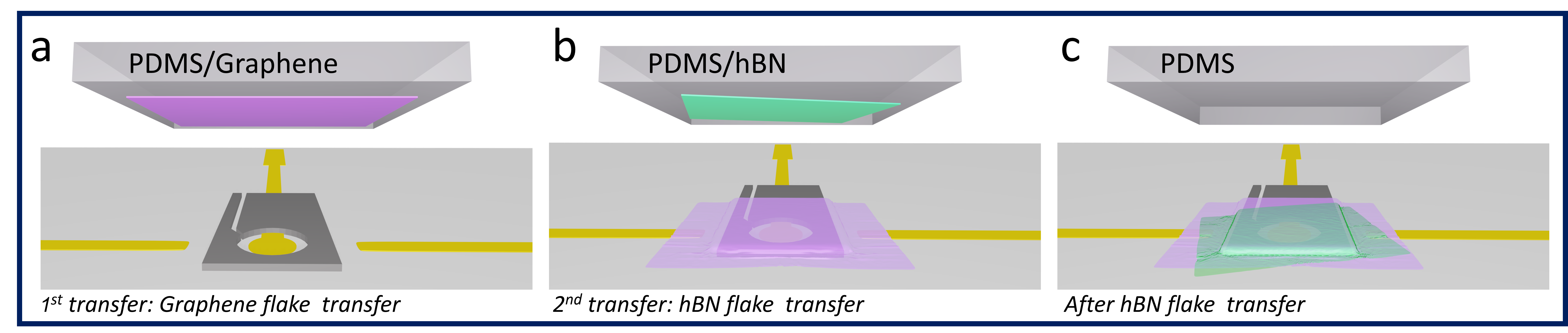}
\caption{\textbf{Schematics of PDMS assisted dry transfer of graphene and hBN on a drum structure.} (a) Graphene transfer onto the HSQ drum hole. (b) hBN transfer over graphene on the HSQ drum hole. (c) Whole drum assembly after graphene and hBN transfer.}
\label{fig:Supp Dry transfer flow}
\end{figure}

Figure \ref{fig:Supp Dry transfer flow} shows the schematic of graphene and hBN transfer. We use polydimethylsiloxane (PDMS) assisted dry transfer method\cite{SI_Castellanos-Gomez2014DeterministicStamping}. We exfoliated the graphene and the hBN on different PDMS stamps and transferred one after another onto the drum structure at room temperature. Graphene flake is transferred  first (Figure \ref{fig:Supp Dry transfer flow}a) and then hBN (Figure \ref{fig:Supp Dry transfer flow}b). Both graphene and hBN thicknesses are $\approx 10$~nm.

\subsection{Device fabrication} 

The devices are fabricated on Sapphire substrate to reduce the effect of parasitic capacitance. Source, drain, and local gate of Cr/Au (5 and 50~nm) electrodes are deposited on the substrate. The drum structure is fabricated on the gate electrode with hydrogen silsesquioxane (HSQ) e-beam negative resist. The diameter and height of the drum holes are 3.5 {\textmu}m and 250~nm respectively, and the diameter of the gate electrode is  2 {\textmu}m. The drum resonators D1, D2 and D3 are fabricated by transferring first a mechanically exfoliated few-layer graphene and then a few-layer hBN on top of it to the pre-patterned holes using PDMS dry transfer technique\cite{SI_Castellanos-Gomez2014DeterministicStamping} (Figure \ref{fig:Supp Dry transfer flow}). 

\subsection{Measurements} 

Optical detection measurement is carried out using a 633~nm laser with a power of $\approx$150~\textmu W incident at the device. We use 20x objective and photodiode (New Focus 1801) for detection. Mode shape mappings were obtained by scanning the device using XY piezo stage.


\section{Room temperature data of D1, D2 and D3}

\begin{figure}[h]
\centering
\includegraphics[scale=0.8]{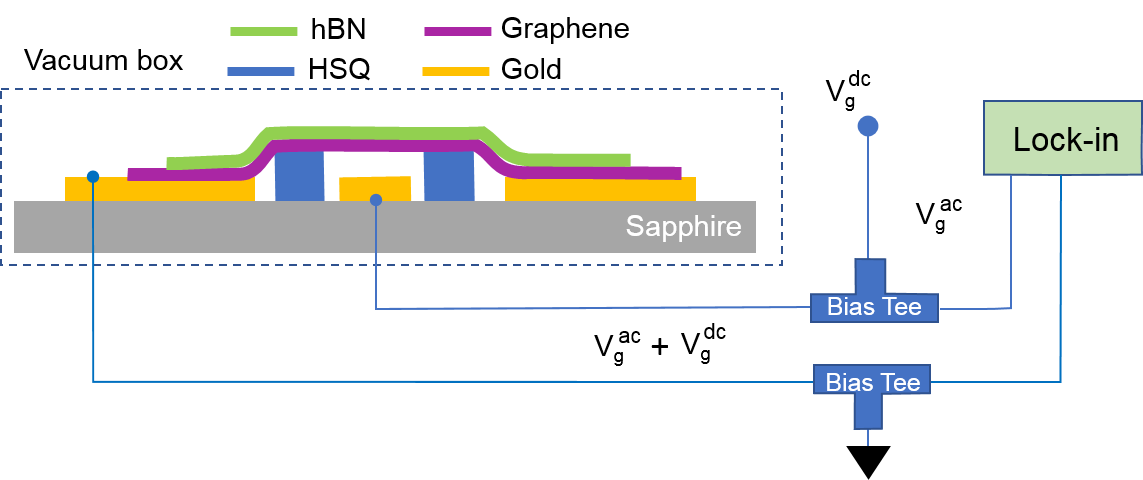}
\caption{\textbf{Electrical detection scheme.}  Schematic diagrams of the drum resonator cross-section under electrical actuation and electrical detection measurement scheme.}
\label{fig:Supp electriacl detection scheme}
\end{figure}

\begin{figure}[h]
\centering
\includegraphics[scale=1]{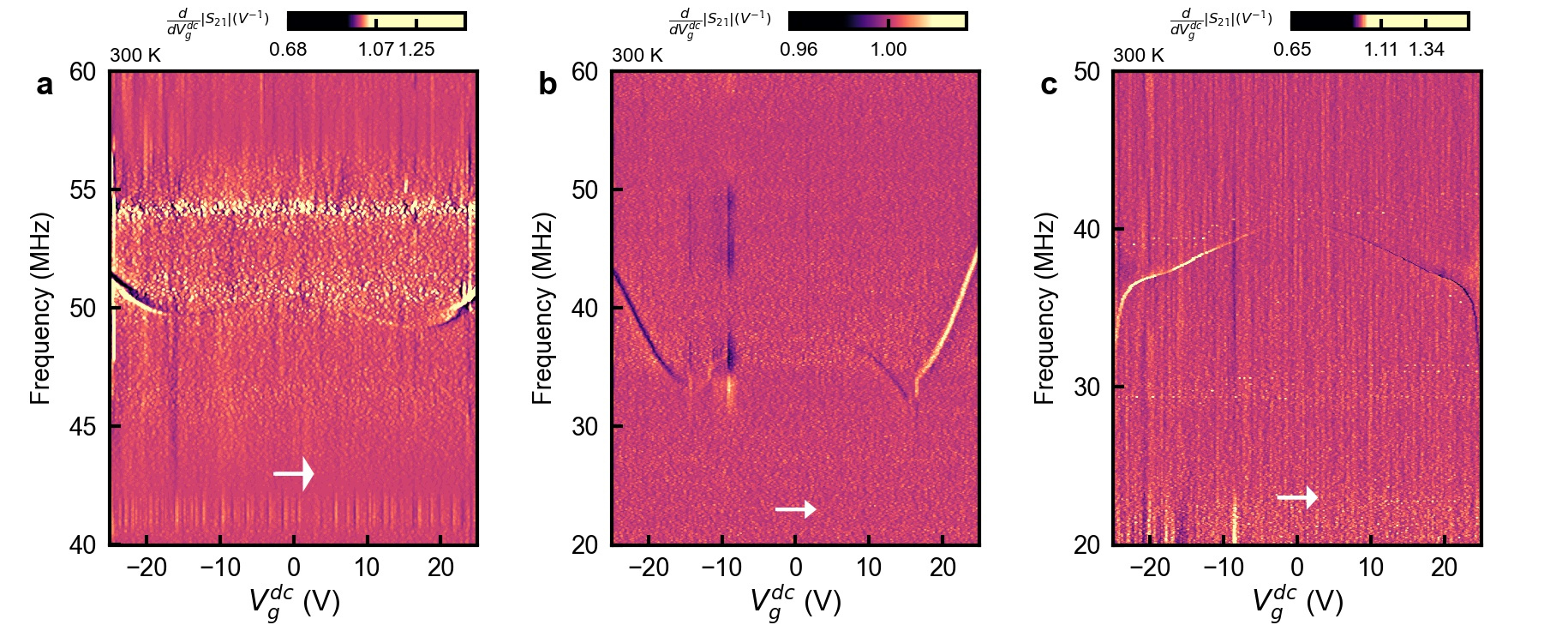}
\caption{\textbf{Room temperature data.}:a,b and c)Color plots of D1 , D2 and D3 device frequency dispersion curves as function of gate voltage respectively.The white arrows indicate the gate voltage sweep direction}
\label{fig:Room temperature data}
\end{figure}

Figure \ref{fig:Room temperature data}a,b and c show  room temperature frequency dispersion color plots of D1, D2 and D3 respectively. They are measured using the electrical detection scheme, shown in Figure \ref{fig:Supp electriacl detection scheme}.

\section{Initial asymmetry in the device D1 and D2}

\begin{figure}[h]
\centering
\includegraphics[scale=1]{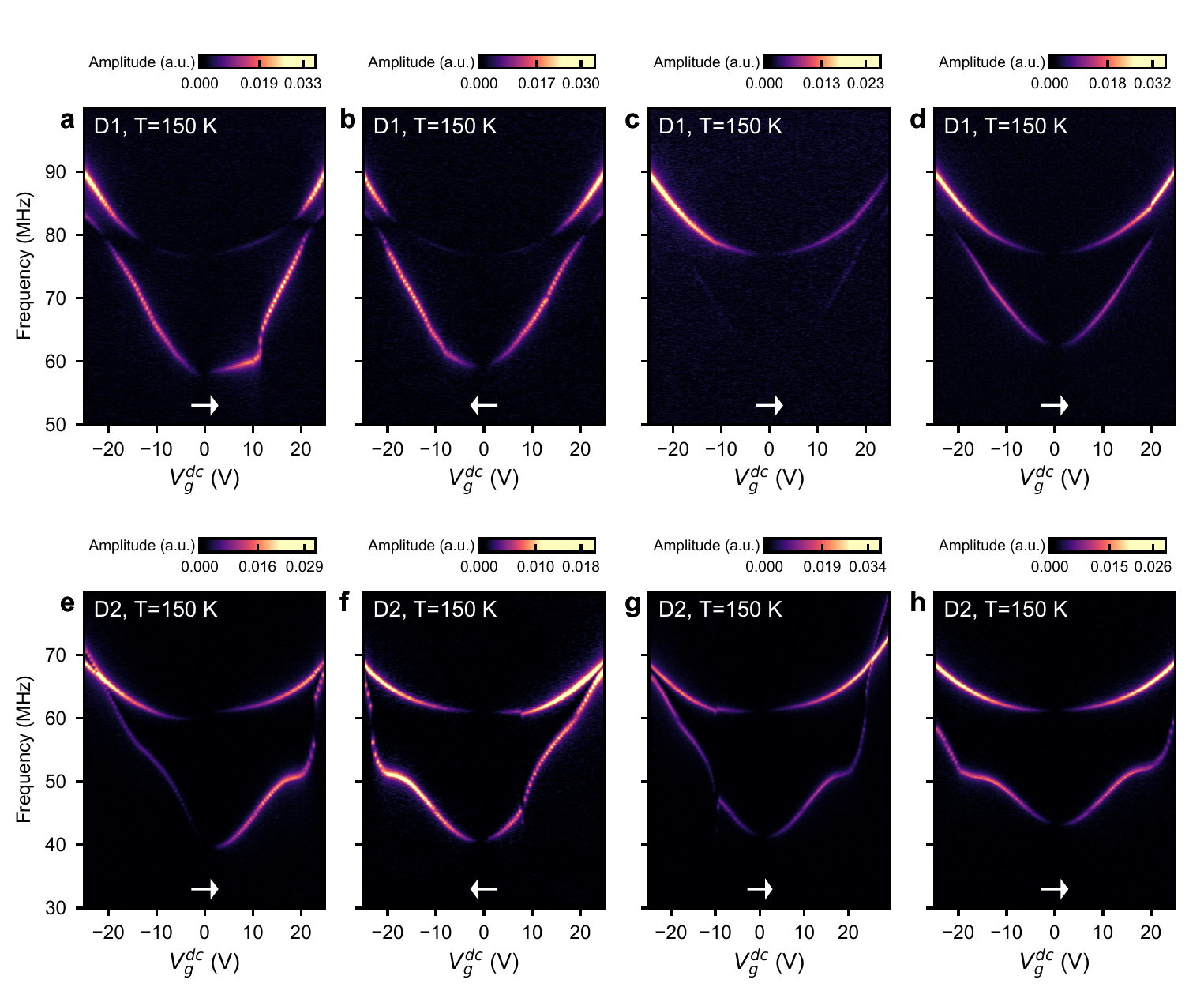}
\caption{\textbf{Asymmetry in initial measurement cycles in D1 and D2.} (a,b,c,\&d) 1$^\textrm{st}$, 2$^\textrm{nd}$, 3$^\textrm{rd}$, and 4$^\textrm{th}$ frequency sweeps of the device D1. (e,f,g,\&h) 1$^\textrm{st}$, 2$^\textrm{nd}$, 3$^\textrm{rd}$, and 4$^\textrm{th}$ frequency sweeps of the device D2. The color plot represents the reflected laser signal proportional to the amplitude of the membrane motion. The white arrows indicate the direction of the gate voltage sweep.}
\label{fig:Supp D1&D2 few cycles}
\end{figure}

Figure \ref{fig:Supp D1&D2 few cycles} shows the frequency dispersion curves for the first four cycles of measurement on devices D1 and D2. They showed asymmetry in the frequency response around the neutrality point ($V_g^\textrm{dc}=0$~V) for the first few cycles. The asymmetry reduced for every next sweep and after almost four sweeps, both D1 and D2 showed symmetric behaviour with gate voltage without any changes in further sweeps (Figure \ref{fig:Supp D1&D2 few cycles}d and h). This asymmetry could arise from unstable strains induced in the transfer process that gradually equilibrates through subsequent sweeps of gate voltage. No such stain relaxation was observed in device D3. Repeatability of measured features and stability of bubble morphology over months show the impermeable nature of the interface\cite{SI_Sun2020LimitsGraphene}, and a stable in-built tension in the membrane.

\section{New mode at higher frequency}
\begin{figure}[h]
\centering
\includegraphics[scale=1.5]{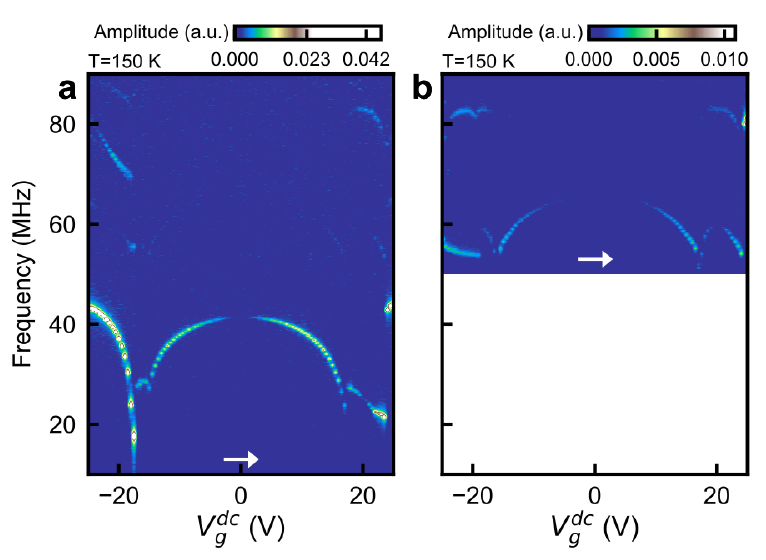}
\caption{\textbf{Broad range frequency scan for the device D3 at} $T=$150~K. a) Frequency scan of D3 ranging from 10~MHz to 100~MHz with RF power $-30$~dBm. b) Frequency scan of D3 ranging from 50~MHz to 100~MHz with RF power $-20$~dBm. White arrows indicate the direction of the sweep}
\label{fig:Supp Higher modes of D3}
\end{figure}

From Figure \ref{fig:Supp Higher modes of D3}a, a new mode with a higher resonance frequency ($\approx$ 75~MHz) is observed only in the region $R3^\prime$ in a broader frequency scan window. A notable aspect of this new mode is that it does not exist in the regions $R2^\prime$ and $R3^\prime$, and also in the absence of the primary mode (between 20 to 40~MHz), shown in Figure \ref{fig:Supp Higher modes of D3}b. This suggests that the new mode results from the detachment of graphene and hBN in the region $R3^\prime$. This mode might be the resonant mode of hBN after detaching from graphene, and further supports the proposed interface dynamics..  

\section{Frequency scans of D2 and D3 at different regions of $V_g^\textrm{dc}$}
\begin{figure}[h]
\centering
\includegraphics[scale=1]{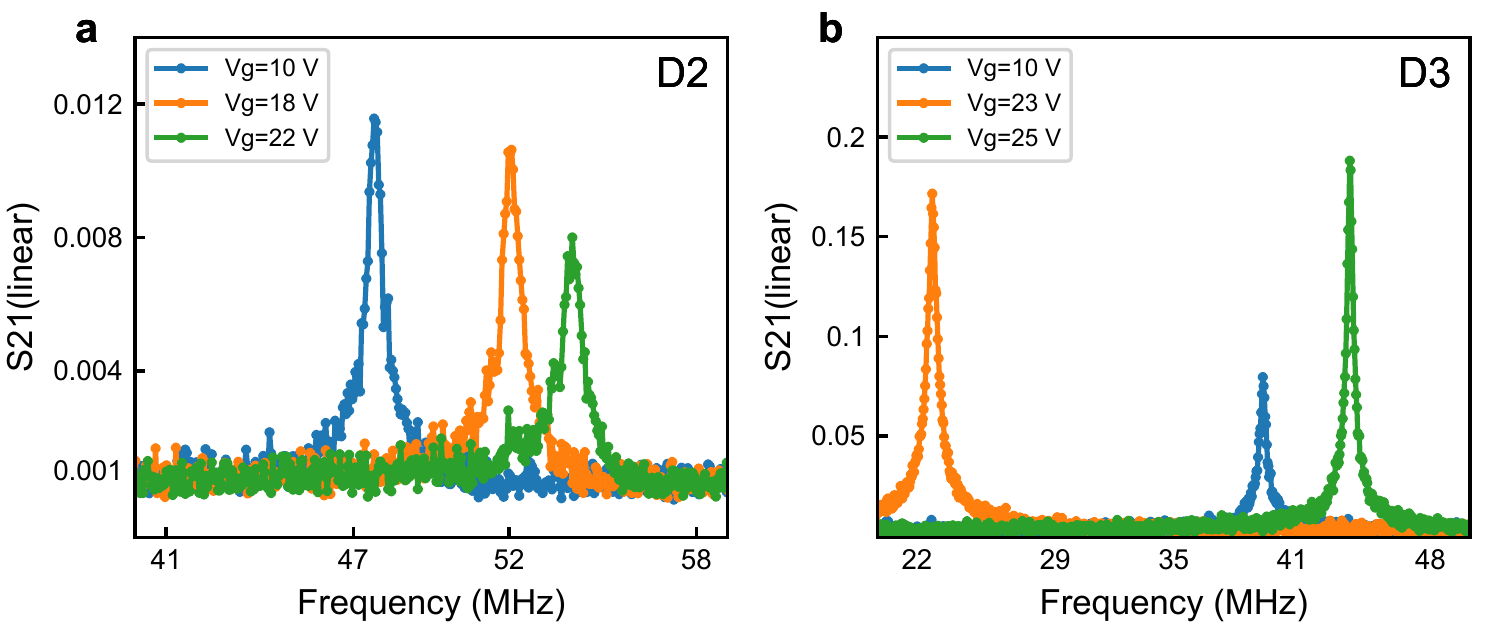}
\caption{\textbf{Frequency scans of D2 and D3 at different gate voltages.} a) Frequency scans of D2 at $V_g^\textrm{dc}=$10~V, 18~V and 22~V. b) Frequency scans of D3 in the region R1 at $V_g^\textrm{dc}=$10~V, in the region R2 at $V_g^\textrm{dc}=$23~V, and the region R3 at $V_g^\textrm{dc}=$25~V.}
\label{fig:Supp D2 and D3 line slices}
\end{figure}

Figure \ref{fig:Supp D2 and D3 line slices} shows the frequency scans in regions of different $V_g^\textrm{dc}$ for the devices D2 and D3 indicated in the main text, Figure 2a\&b, respectively. These line scans show that the devices D2 and D3 are in linear regime at all $V_g^\textrm{dc}$ values.

\section{Hysteresis  in the device D2 response}
\begin{figure}[h]
\centering
\includegraphics[scale=0.9]{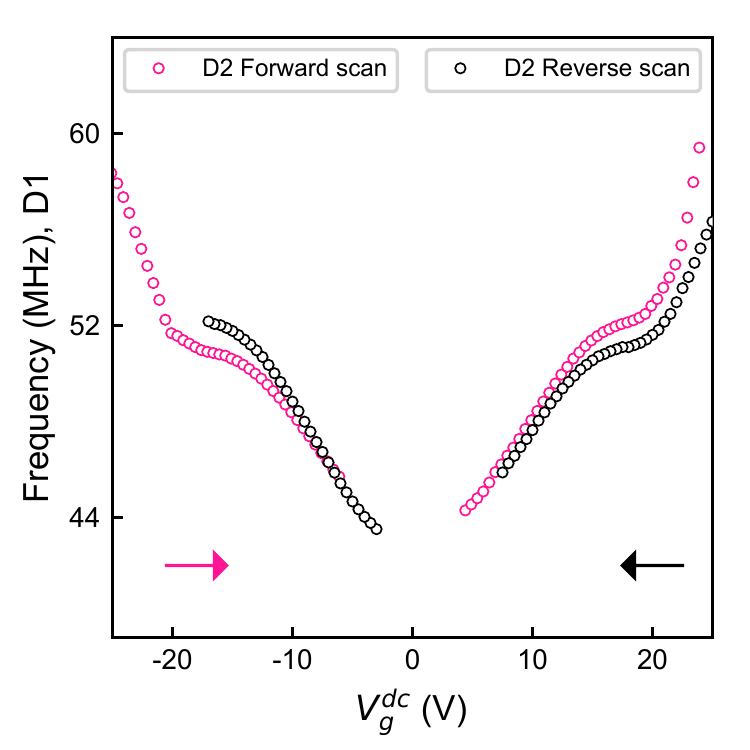}
\caption{\textbf{Hysteresis in the frequency dispersion of D2.} Frequency dispersion of fundamental mode in the forward direction from $-25$~V to 25~V (pink curve) and the reverse direction (black curve) shows clear presence of a hysteresis with the direction of sweep of the gate voltage. The white arrows indicate the sweep direction}
\label{fig:Supp D2 hysteresis}
\end{figure}

Figure \ref{fig:Supp D2 hysteresis} shows the hysteresis in device D2 with the direction of sweep of the gate voltage $V_g^\textrm{dc}$. Hysteresis shows up from the kink region onwards ($V_g^\textrm{dc}\geq $14~V). Specifically, in the hysteresis region, the frequency values are relatively higher when the sweep is from high to low gate voltage than in the reverse direction. But the tunability is unchanged with the sweep direction. A small shift in the neutrality point (NP) $\approx$~0.5~V is observed in the device D1. This shift in NP is well known in electrical measurements \cite{SI_Kim2018Nano-electromechanicalBimorphs}.  We also observed  the shift in other devices D1 ($\approx$ 1~V) and D3 ($\approx$ 1~V).

\section{Measurements using electrical detection}

\begin{figure}[h]
\centering
\includegraphics[scale=1.4]{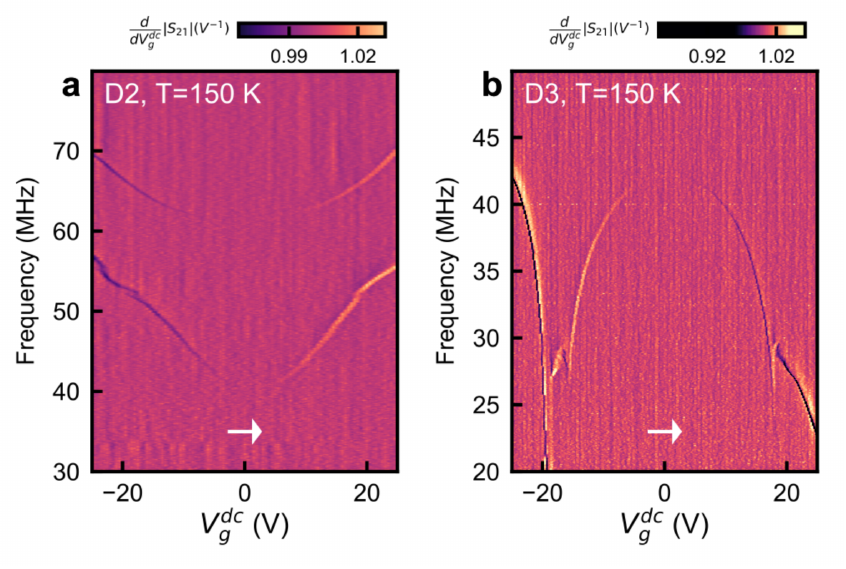}
\caption{\textbf{Frequency dispersion measured using electrical detection} (a,b) Frequency dispersion of D2 and D3 at $T=150$~K respectively using electrical detection scheme (Figure \ref{fig:Supp electriacl detection scheme}). The white arrows indicate the scan direction.}
\label{fig:Supp D2 and D3 electrical detection}
\end{figure}

Figure \ref{fig:Supp D2 and D3 electrical detection} shows  the frequency dispersion of D2 and D3 measured using the electrical detection scheme at T=150 K, as shown in Figure \ref{fig:Supp electriacl detection scheme}. We fabricate our devices on a sapphire substrate with a local back gate to reduce the parasitic capacitance in an electrical detection scheme. The frequency response using electrical detection shows identical to that observed using an optical scheme (Figure 1e\&f). So we can rule out the heating effect due to the laser in the devices D2 and D3 response.


\section{Additional device data  including graphene on hBN}

\begin{figure}[h]
\centering
\includegraphics[scale=0.7]{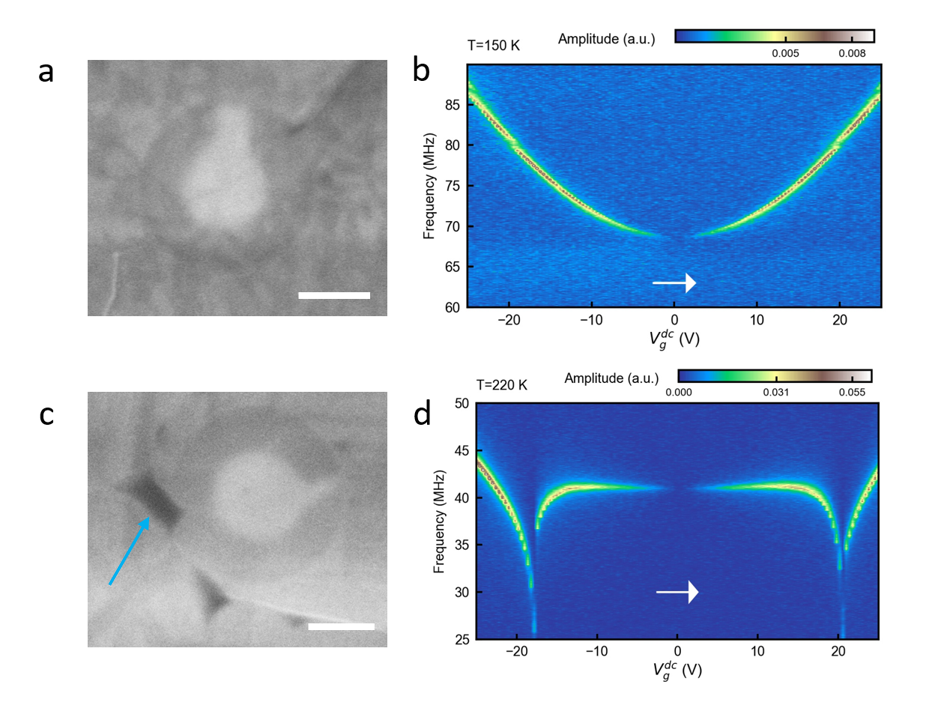}
\caption{\textbf{Variation of the resonant frequency for additional bubble-free drum and a drum with bubble }: a) and c) SEM images on hBN on graphene drums with clean and bubble interfaces respectively. Scale bar in the SEM images is equal to 2 µm. Black color regions in SEM image (panel c) are the bubble regions at the heterostructure interface and it is also indicated by blue arrow. b) and d) The color-scale plots show the variation of the resonant frequency as a function of the gate voltage for a drum without bubble (panel a) and with bubble (panel c) respectively . White arrow in the color plots indicates the gate sweep direction. This data is similar to the response of D1 (figure 1b and 1e) and D3 (Figure 1d and 1g) in the main manuscript.}
\label{fig:More device data}
\end{figure}

We have made additional devices with different cleanliness  at the interface of the graphene-hBN heterostructure. Figure \ref{fig:More device data}a and c show the SEM images of the drums with clean interface and with bubble respectively. Their frequency dispersion as a function of gate voltage is shown in Figure \ref{fig:More device data}b and d respectively. This data is similar to the response of D1 (figure 1b and 1e) and D3 (figure 1d and 1g) in the main manuscript. 

\begin{figure}[h]
\centering
\includegraphics[scale=0.5]{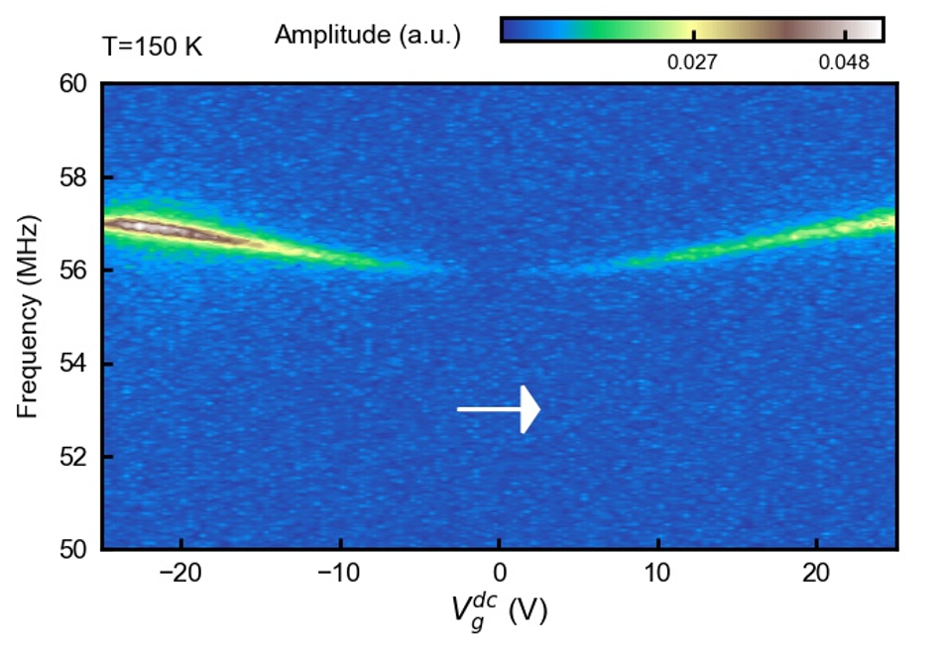}
\caption{\textbf{Graphene on hBN drum response} : Frequency response of a graphene on hBN drum at temperature T=150 K, white arrow in the color plot indicates Vg sweep direction which is from -25 V to 25 V. }
\label{fig:BNG drum}
\end{figure}

We have made drums with a reverse sequence of the heterostructure that is graphene on hBN. Its frequency response as function of gate voltage is shown in Figure \ref{fig:BNG drum}. It shows positive dispersion with low tunability of 0.04 MHz/V, which is smaller than hBN on graphene drums. It does not show any kinks or jumps. 

\section{Finite element simulations (FEM)}

\subsection{Simulations} 

We use COMSOL Multiphysics for our FEM simulations. For the bubble growth rate equation, we considered linear gate-voltage dependence. Bubble growth will be restricted as the bubble diameter reaches the diameter of the drum. So the growth is conditional, it increases as the gate voltage increases with a constant slope till its radius reaches the drum radius ($R_f$); its governing equation is $R_{\textrm{bub}}(V_g^{\textrm{dc}})=R_i+mV_g^{\textrm{dc}}$ for $R_{bub}\leq R_f$. Bubble growth stops at $R_{\textrm{bub}} \approx R_f$, where $R_i$, $R_f$ are the initial and final radii of the bubble respectively and $m$ is an increase in bubble radius per unit change in gate voltage (\textmu  m/V). While a linear bubble growth rate is a simplifying assumption to compare with experimental data, it is likely that the actual growth rate is non-linear.

The parameters we used for our simulations are Young's modulus for both graphene and hBN is $ 1\times 10^{12}$~Pa. Density of graphene and hBN are 1950 Kg/m$^3$ and 2100 Kg/m$^3$ respectively \cite{SI_Wang2019BendingMaterials,SI_falin2017mechanical}. The COMSOL codes are publicly accessible via Zenodo (link: https://doi.org/10.5281/zenodo.5929530). 

\subsection{$K_\textrm{vdW}^A$ optimisation using FEM}

\begin{figure}[h]
\centering
\includegraphics[scale=0.5]{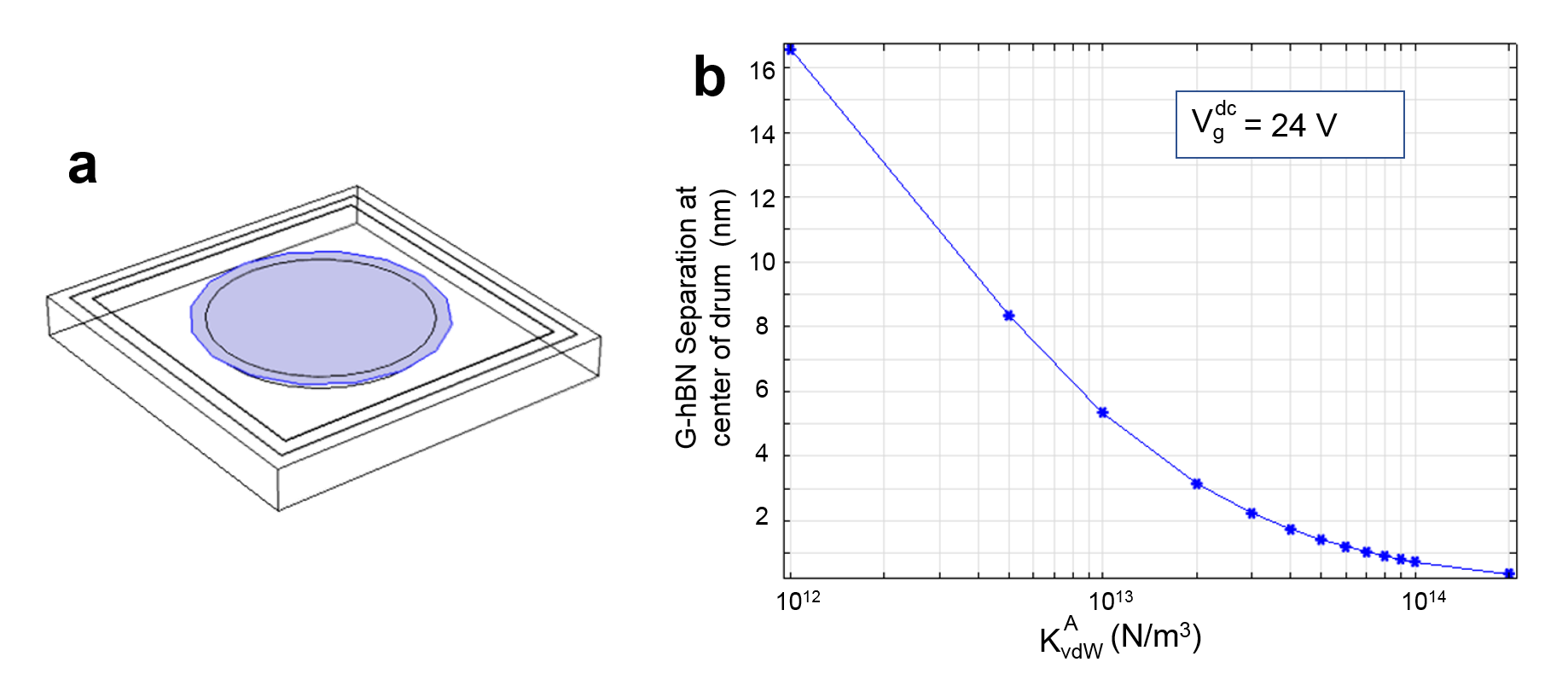}
\caption{\textbf{$K_\textrm{vdW}^A$ optimisation.} a) G-hBN drum geometry for FEM simulation. The blue color region represents a thin-elastic region with a spring constant per unit area $K_\textrm{vdW}^A$. b) Maximum separation between graphene and hBN for different values of $K_\textrm{vdW}^A$.}
\label{fig:Supp Kbub optimisation}
\end{figure}

Figure \ref{fig:Supp Kbub optimisation}a shows the geometry used for finite element modelling (FEM). We used the G-hBN drum without any bubbles to optimize the spring constant per unit area, $K_\textrm{vdW}^A$, due to van der Waals forces. The blue region in the geometry represents the thin elastic region with spring constant per unit area $K_\textrm{vdW}^A$. From the device D1 response shown in Figure 2a, no signature of slip at the G-hBN interface is observed in the absence of bubbles. Hence, the separation between graphene and hBN should be in the order of van der Waals distance less than 1~nm. Figure \ref{fig:Supp Kbub optimisation} shows the separation between graphene and hBN calculated using FEM at different $K_\textrm{vdW}^A$ values. From the simulation results, $K_\textrm{vdW}^A$ should be $\leq 8\times 10^{13}$~N/m$^3$ in the absence or presence of bubbles in the interface. 

\subsection{Stress calculation at the heterostructure interface}
\begin{figure}[h]
\centering
\includegraphics[scale=0.6]{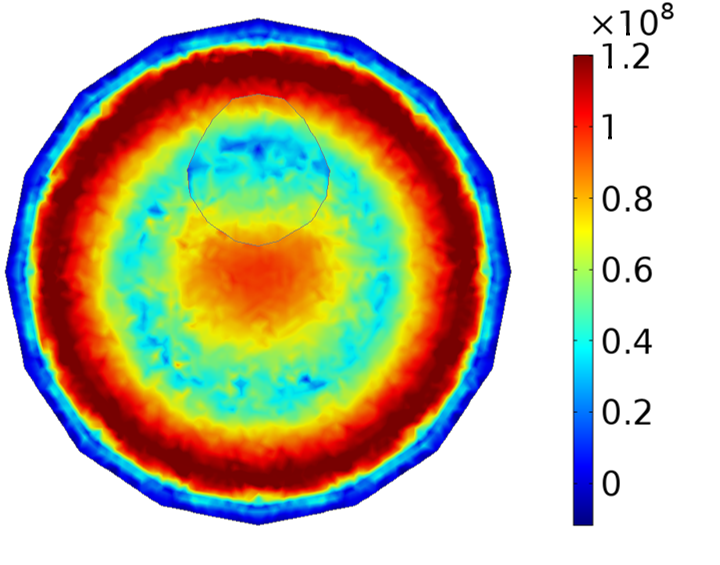}
\caption{\textbf{Stresses at the bubble edges}: The color-scale (unit for color-scale on the right is N/m$^2$) plot shows the stress at the edges of a bubble at 24 V gate voltage.}
\label{fig:Stresses at the bubble edges}
\end{figure}

Figure \ref{fig:Stresses at the bubble edges} shows the simulations in the static geometry to get a quantitative idea about the stress near the bubble at gate voltege of 24~V. From the calculation, the bubble edges show higher stress with boundary stress at the bubble of $\approx 10^8$~N/$m^2$ (Figure \ref{fig:Stresses at the bubble edges}).

\section{Griffith’s picture of fracture}

\begin{figure}[h]
\centering
\includegraphics[scale=0.75]{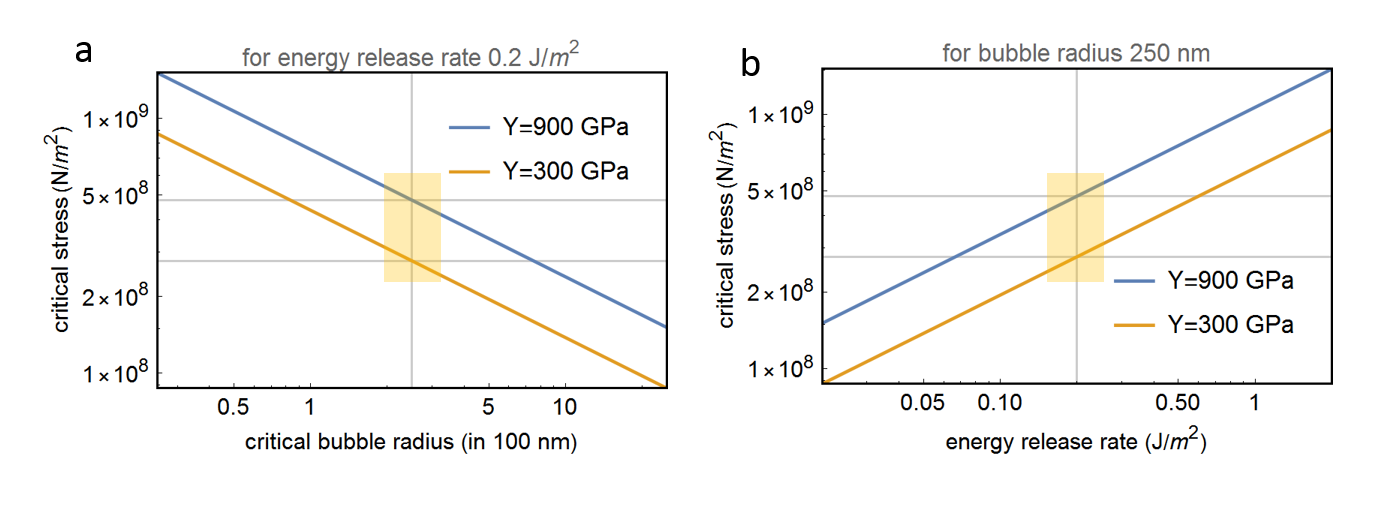}

\caption{\textbf{Critical stress as a function of the energy release rate and critical bubble radius:} a)Plot of Griffith’s critical fracture stress as a function of the bubble dimension. b)  Plot of Griffith’s critical fracture stress as a function of the energy release rate for a given bubble dimension (radius = 250 nm). The yellow shaded region indicates the parameter space where we see delamination. Two values of Young’s modulus are used to indicate that even after accounting for the reduction of bending stiffness we see results that are consistent across the experiments and fracture picture. }
\label{fig:critical stress}
\end{figure}

Now using Griffith’s simple picture $\sigma _F=\sqrt{(G_cY)/(\pi a)}$; here $\sigma_F$ is the fracture stress and $G_c$
is the energy release rate, analogous to adhesion energy,  $Y$ is the Young’s modulus and $2a$ is the critical fracture length ($a$ is the radius of the bubble). 

Now, we use the adhesion energy reported in the literature and show a critical fracture length. While adhesion energy estimates vary in the literature it is $\approx$ 0.2~J/$m^2$ \cite{SI_dai2020mechanicsopportunities}. If we use these values as $G_c$ we can understand our experimental observations. 

The static stress profile that we see at the edge of the bubble is $\approx ~10^8$ N/$m^2$. As we see, this stress is very close to critical stress needed for fracture for large bubbles. Using the values known in the literature for Y \cite{SI_Wang2019BendingMaterials} we find the critical fracture length to be 500 nm (corresponding to bubble radius of 250 nm). This is consistent with the dimension of the bubbles where we see delamination.

Using the literature adhesion energy of about 0.2 J/m$^2$ we plot the curve for the critical stress with varying bubble radius, shown in Figure \ref{fig:critical stress}a and we find that at gate voltages of $\approx$~24 V we reach a critical stress value of $\approx ~10^8$ N/$m^2$ seen from our COMSOL simulations for bubble radius of $\approx$~250 nm (with a diameter 500 nm). From the Figure \ref{fig:critical stress}a, it indicates reasonable agreement with our experimental observations. 

Figure \ref{fig:critical stress}b shows the plot of Griffith’s critical fracture stress as a function of the energy release rate for a given bubble dimension (radius = 250 nm), Which is close to our bubble size. Combining the analysis and the experimental results we see a critical bubble radius of $\approx$~250 nm; implying a critical diameter of 500 nm. Our results are also consistent with reported adhesion energies.


\begin{thebibliography}{52}%
\makeatletter
\providecommand \@ifxundefined [1]{%
 \@ifx{#1\undefined}
}%
\providecommand \@ifnum [1]{%
 \ifnum #1\expandafter \@firstoftwo
 \else \expandafter \@secondoftwo
 \fi
}%
\providecommand \@ifx [1]{%
 \ifx #1\expandafter \@firstoftwo
 \else \expandafter \@secondoftwo
 \fi
}%
\providecommand \natexlab [1]{#1}%
\providecommand \enquote  [1]{``#1''}%
\providecommand \bibnamefont  [1]{#1}%
\providecommand \bibfnamefont [1]{#1}%
\providecommand \citenamefont [1]{#1}%
\providecommand \href@noop [0]{\@secondoftwo}%
\providecommand \href [0]{\begingroup \@sanitize@url \@href}%
\providecommand \@href[1]{\@@startlink{#1}\@@href}%
\providecommand \@@href[1]{\endgroup#1\@@endlink}%
\providecommand \@sanitize@url [0]{\catcode `\\12\catcode `\$12\catcode
  `\&12\catcode `\#12\catcode `\^12\catcode `\_12\catcode `\%12\relax}%
\providecommand \@@startlink[1]{}%
\providecommand \@@endlink[0]{}%
\providecommand \url  [0]{\begingroup\@sanitize@url \@url }%
\providecommand \@url [1]{\endgroup\@href {#1}{\urlprefix }}%
\providecommand \urlprefix  [0]{URL }%
\providecommand \Eprint [0]{\href }%
\providecommand \doibase [0]{http://dx.doi.org/}%
\providecommand \selectlanguage [0]{\@gobble}%
\providecommand \bibinfo  [0]{\@secondoftwo}%
\providecommand \bibfield  [0]{\@secondoftwo}%
\providecommand \translation [1]{[#1]}%
\providecommand \BibitemOpen [0]{}%
\providecommand \bibitemStop [0]{}%
\providecommand \bibitemNoStop [0]{.\EOS\space}%
\providecommand \EOS [0]{\spacefactor3000\relax}%
\providecommand \BibitemShut  [1]{\csname bibitem#1\endcsname}%
\let\auto@bib@innerbib\@empty
\bibitem [{\citenamefont {Lee}\ \emph {et~al.}(2008)\citenamefont {Lee},
  \citenamefont {Wei}, \citenamefont {Kysar},\ and\ \citenamefont
  {Hone}}]{Lee2008MeasurementGraphene}%
  \BibitemOpen
  \bibfield  {author} {\bibinfo {author} {\bibfnamefont {C.}~\bibnamefont
  {Lee}}, \bibinfo {author} {\bibfnamefont {X.}~\bibnamefont {Wei}}, \bibinfo
  {author} {\bibfnamefont {J.~W.}\ \bibnamefont {Kysar}}, \ and\ \bibinfo
  {author} {\bibfnamefont {J.}~\bibnamefont {Hone}},\ }\href {\doibase
  10.1126/science.1157996} {\bibfield  {journal} {\bibinfo  {journal}
  {Science}\ }\textbf {\bibinfo {volume} {321}},\ \bibinfo {pages} {385}
  (\bibinfo {year} {2008})}\BibitemShut {NoStop}%
\bibitem [{\citenamefont {Yu}\ \emph {et~al.}(2021{\natexlab{a}})\citenamefont
  {Yu}, \citenamefont {Hossain}, \citenamefont {Kim}, \citenamefont {Ferrari},
  \citenamefont {Huang}, \citenamefont {Zhang}, \citenamefont {Kim},
  \citenamefont {Michel},\ and\ \citenamefont {{van der
  Zande}}}]{Yu2021MechanicallyMembranes}%
  \BibitemOpen
  \bibfield  {author} {\bibinfo {author} {\bibfnamefont {J.}~\bibnamefont
  {Yu}}, \bibinfo {author} {\bibfnamefont {M.~A.}\ \bibnamefont {Hossain}},
  \bibinfo {author} {\bibfnamefont {S.}~\bibnamefont {Kim}}, \bibinfo {author}
  {\bibfnamefont {P.~F.}\ \bibnamefont {Ferrari}}, \bibinfo {author}
  {\bibfnamefont {S.}~\bibnamefont {Huang}}, \bibinfo {author} {\bibfnamefont
  {Y.}~\bibnamefont {Zhang}}, \bibinfo {author} {\bibfnamefont
  {H.}~\bibnamefont {Kim}}, \bibinfo {author} {\bibfnamefont {D.~A.}\
  \bibnamefont {Michel}}, \ and\ \bibinfo {author} {\bibfnamefont {A.~M.}\
  \bibnamefont {{van der Zande}}},\ }\href {\doibase
  https://doi.org/10.1016/j.cossms.2021.100900} {\bibfield  {journal} {\bibinfo
   {journal} {Current Opinion in Solid State and Materials Science}\ }\textbf
  {\bibinfo {volume} {25}},\ \bibinfo {pages} {100900,
  10.1016/j.cossms.2021.100900} (\bibinfo {year}
  {2021}{\natexlab{a}})}\BibitemShut {NoStop}%
\bibitem [{\citenamefont {Bunch}\ \emph {et~al.}(2007)\citenamefont {Bunch},
  \citenamefont {van~der Zande}, \citenamefont {Verbridge}, \citenamefont
  {Frank}, \citenamefont {Tanenbaum}, \citenamefont {Parpia}, \citenamefont
  {Craighead},\ and\ \citenamefont
  {McEuen}}]{Bunch2007ElectromechanicalSheets}%
  \BibitemOpen
  \bibfield  {author} {\bibinfo {author} {\bibfnamefont {J.~S.}\ \bibnamefont
  {Bunch}}, \bibinfo {author} {\bibfnamefont {A.~M.}\ \bibnamefont {van~der
  Zande}}, \bibinfo {author} {\bibfnamefont {S.~S.}\ \bibnamefont {Verbridge}},
  \bibinfo {author} {\bibfnamefont {I.~W.}\ \bibnamefont {Frank}}, \bibinfo
  {author} {\bibfnamefont {D.~M.}\ \bibnamefont {Tanenbaum}}, \bibinfo {author}
  {\bibfnamefont {J.~M.}\ \bibnamefont {Parpia}}, \bibinfo {author}
  {\bibfnamefont {H.~G.}\ \bibnamefont {Craighead}}, \ and\ \bibinfo {author}
  {\bibfnamefont {P.~L.}\ \bibnamefont {McEuen}},\ }\href {\doibase
  10.1126/science.1136836} {\bibfield  {journal} {\bibinfo  {journal}
  {Science}\ }\textbf {\bibinfo {volume} {315}},\ \bibinfo {pages} {490}
  (\bibinfo {year} {2007})}\BibitemShut {NoStop}%
\bibitem [{\citenamefont {Mathew}\ \emph {et~al.}(2016)\citenamefont {Mathew},
  \citenamefont {Patel}, \citenamefont {Borah}, \citenamefont {Vijay},\ and\
  \citenamefont {Deshmukh}}]{Mathew2016DynamicalDrums}%
  \BibitemOpen
  \bibfield  {author} {\bibinfo {author} {\bibfnamefont {J.~P.}\ \bibnamefont
  {Mathew}}, \bibinfo {author} {\bibfnamefont {R.~N.}\ \bibnamefont {Patel}},
  \bibinfo {author} {\bibfnamefont {A.}~\bibnamefont {Borah}}, \bibinfo
  {author} {\bibfnamefont {R.}~\bibnamefont {Vijay}}, \ and\ \bibinfo {author}
  {\bibfnamefont {M.~M.}\ \bibnamefont {Deshmukh}},\ }\href {\doibase
  10.1038/nnano.2016.94} {\bibfield  {journal} {\bibinfo  {journal} {Nature
  Nanotechnology}\ }\textbf {\bibinfo {volume} {11}},\ \bibinfo {pages} {747}
  (\bibinfo {year} {2016})}\BibitemShut {NoStop}%
\bibitem [{\citenamefont {Zheng}\ \emph {et~al.}(2017)\citenamefont {Zheng},
  \citenamefont {Lee},\ and\ \citenamefont {Feng}}]{Zheng2017HexagonalMotion}%
  \BibitemOpen
  \bibfield  {author} {\bibinfo {author} {\bibfnamefont {X.-Q.}\ \bibnamefont
  {Zheng}}, \bibinfo {author} {\bibfnamefont {J.}~\bibnamefont {Lee}}, \ and\
  \bibinfo {author} {\bibfnamefont {P.~X.-L.}\ \bibnamefont {Feng}},\ }\href
  {\doibase 10.1038/micronano.2017.38} {\bibfield  {journal} {\bibinfo
  {journal} {Microsystems {\&} Nanoengineering}\ }\textbf {\bibinfo {volume}
  {3}},\ \bibinfo {pages} {17038} (\bibinfo {year} {2017})}\BibitemShut
  {NoStop}%
\bibitem [{\citenamefont {Lee}\ \emph {et~al.}(2018)\citenamefont {Lee},
  \citenamefont {Wang}, \citenamefont {He}, \citenamefont {Yang}, \citenamefont
  {Shan},\ and\ \citenamefont {Feng}}]{Lee2018ElectricallyRange}%
  \BibitemOpen
  \bibfield  {author} {\bibinfo {author} {\bibfnamefont {J.}~\bibnamefont
  {Lee}}, \bibinfo {author} {\bibfnamefont {Z.}~\bibnamefont {Wang}}, \bibinfo
  {author} {\bibfnamefont {K.}~\bibnamefont {He}}, \bibinfo {author}
  {\bibfnamefont {R.}~\bibnamefont {Yang}}, \bibinfo {author} {\bibfnamefont
  {J.}~\bibnamefont {Shan}}, \ and\ \bibinfo {author} {\bibfnamefont {P.~X.}\
  \bibnamefont {Feng}},\ }\href {\doibase 10.1126/sciadv.aao6653} {\bibfield
  {journal} {\bibinfo  {journal} {Science Advances}\ }\textbf {\bibinfo
  {volume} {4}},\ \bibinfo {pages} {eaao6653, 10.1126/sciadv.aao6653} (\bibinfo
  {year} {2018})}\BibitemShut {NoStop}%
\bibitem [{\citenamefont {Jiang}\ \emph {et~al.}(2020)\citenamefont {Jiang},
  \citenamefont {Xie}, \citenamefont {Shan},\ and\ \citenamefont
  {Mak}}]{Jiang2020ExchangeAntiferromagnets}%
  \BibitemOpen
  \bibfield  {author} {\bibinfo {author} {\bibfnamefont {S.}~\bibnamefont
  {Jiang}}, \bibinfo {author} {\bibfnamefont {H.}~\bibnamefont {Xie}}, \bibinfo
  {author} {\bibfnamefont {J.}~\bibnamefont {Shan}}, \ and\ \bibinfo {author}
  {\bibfnamefont {K.~F.}\ \bibnamefont {Mak}},\ }\href {\doibase
  10.1038/s41563-020-0712-x} {\bibfield  {journal} {\bibinfo  {journal} {Nature
  Materials}\ }\textbf {\bibinfo {volume} {19}},\ \bibinfo {pages} {1295}
  (\bibinfo {year} {2020})}\BibitemShut {NoStop}%
\bibitem [{\citenamefont {Koenig}\ \emph {et~al.}(2012)\citenamefont {Koenig},
  \citenamefont {Wang}, \citenamefont {Pellegrino},\ and\ \citenamefont
  {Bunch}}]{Koenig2012SelectiveGraphene}%
  \BibitemOpen
  \bibfield  {author} {\bibinfo {author} {\bibfnamefont {S.~P.}\ \bibnamefont
  {Koenig}}, \bibinfo {author} {\bibfnamefont {L.}~\bibnamefont {Wang}},
  \bibinfo {author} {\bibfnamefont {J.}~\bibnamefont {Pellegrino}}, \ and\
  \bibinfo {author} {\bibfnamefont {J.~S.}\ \bibnamefont {Bunch}},\ }\href
  {\doibase 10.1038/nnano.2012.162} {\bibfield  {journal} {\bibinfo  {journal}
  {Nature Nanotechnology}\ }\textbf {\bibinfo {volume} {7}},\ \bibinfo {pages}
  {728} (\bibinfo {year} {2012})}\BibitemShut {NoStop}%
\bibitem [{\citenamefont {Patel}\ \emph {et~al.}(2016)\citenamefont {Patel},
  \citenamefont {Mathew}, \citenamefont {Borah},\ and\ \citenamefont
  {Deshmukh}}]{Patel2016LowSensing}%
  \BibitemOpen
  \bibfield  {author} {\bibinfo {author} {\bibfnamefont {R.~N.}\ \bibnamefont
  {Patel}}, \bibinfo {author} {\bibfnamefont {J.~P.}\ \bibnamefont {Mathew}},
  \bibinfo {author} {\bibfnamefont {A.}~\bibnamefont {Borah}}, \ and\ \bibinfo
  {author} {\bibfnamefont {M.~M.}\ \bibnamefont {Deshmukh}},\ }\href {\doibase
  10.1088/2053-1583/3/1/011003} {\bibfield  {journal} {\bibinfo  {journal} {2D
  Materials}\ }\textbf {\bibinfo {volume} {3}},\ \bibinfo {pages} {011003}
  (\bibinfo {year} {2016})}\BibitemShut {NoStop}%
\bibitem [{\citenamefont {Chaste}\ \emph {et~al.}(2012)\citenamefont {Chaste},
  \citenamefont {Eichler}, \citenamefont {Moser}, \citenamefont {Ceballos},
  \citenamefont {Rurali},\ and\ \citenamefont
  {Bachtold}}]{Chaste2012AResolution}%
  \BibitemOpen
  \bibfield  {author} {\bibinfo {author} {\bibfnamefont {J.}~\bibnamefont
  {Chaste}}, \bibinfo {author} {\bibfnamefont {A.}~\bibnamefont {Eichler}},
  \bibinfo {author} {\bibfnamefont {J.}~\bibnamefont {Moser}}, \bibinfo
  {author} {\bibfnamefont {G.}~\bibnamefont {Ceballos}}, \bibinfo {author}
  {\bibfnamefont {R.}~\bibnamefont {Rurali}}, \ and\ \bibinfo {author}
  {\bibfnamefont {A.}~\bibnamefont {Bachtold}},\ }\href {\doibase
  10.1038/nnano.2012.42} {\bibfield  {journal} {\bibinfo  {journal} {Nature
  Nanotechnology 2012}\ }\textbf {\bibinfo {volume} {7}},\ \bibinfo {pages}
  {301} (\bibinfo {year} {2012})}\BibitemShut {NoStop}%
\bibitem [{\citenamefont {Liu}\ \emph {et~al.}(2014)\citenamefont {Liu},
  \citenamefont {Suk}, \citenamefont {Boddeti}, \citenamefont {Cantley},
  \citenamefont {Wang}, \citenamefont {Gray}, \citenamefont {Hall},
  \citenamefont {Bright}, \citenamefont {Rogers}, \citenamefont {Dunn},
  \citenamefont {Ruoff},\ and\ \citenamefont {Bunch}}]{Liu2014LargeSwitches}%
  \BibitemOpen
  \bibfield  {author} {\bibinfo {author} {\bibfnamefont {X.}~\bibnamefont
  {Liu}}, \bibinfo {author} {\bibfnamefont {J.~W.}\ \bibnamefont {Suk}},
  \bibinfo {author} {\bibfnamefont {N.~G.}\ \bibnamefont {Boddeti}}, \bibinfo
  {author} {\bibfnamefont {L.}~\bibnamefont {Cantley}}, \bibinfo {author}
  {\bibfnamefont {L.}~\bibnamefont {Wang}}, \bibinfo {author} {\bibfnamefont
  {J.~M.}\ \bibnamefont {Gray}}, \bibinfo {author} {\bibfnamefont {H.~J.}\
  \bibnamefont {Hall}}, \bibinfo {author} {\bibfnamefont {V.~M.}\ \bibnamefont
  {Bright}}, \bibinfo {author} {\bibfnamefont {C.~T.}\ \bibnamefont {Rogers}},
  \bibinfo {author} {\bibfnamefont {M.~L.}\ \bibnamefont {Dunn}}, \bibinfo
  {author} {\bibfnamefont {R.~S.}\ \bibnamefont {Ruoff}}, \ and\ \bibinfo
  {author} {\bibfnamefont {J.~S.}\ \bibnamefont {Bunch}},\ }\href {\doibase
  10.1002/adma.201304949} {\bibfield  {journal} {\bibinfo  {journal} {Advanced
  Materials}\ }\textbf {\bibinfo {volume} {26}},\ \bibinfo {pages} {1571}
  (\bibinfo {year} {2014})}\BibitemShut {NoStop}%
\bibitem [{\citenamefont {Geim}\ and\ \citenamefont
  {Grigorieva}(2013)}]{Geim2013VanHeterostructures}%
  \BibitemOpen
  \bibfield  {author} {\bibinfo {author} {\bibfnamefont {A.~K.}\ \bibnamefont
  {Geim}}\ and\ \bibinfo {author} {\bibfnamefont {I.~V.}\ \bibnamefont
  {Grigorieva}},\ }\href {\doibase 10.1038/nature12385} {\bibfield  {journal}
  {\bibinfo  {journal} {Nature 2013}\ }\textbf {\bibinfo {volume} {499}},\
  \bibinfo {pages} {419} (\bibinfo {year} {2013})}\BibitemShut {NoStop}%
\bibitem [{\citenamefont {Watanabe}\ \emph {et~al.}(2004)\citenamefont
  {Watanabe}, \citenamefont {Taniguchi},\ and\ \citenamefont
  {Kanda}}]{Watanabe2004Direct-bandgapCrystal}%
  \BibitemOpen
  \bibfield  {author} {\bibinfo {author} {\bibfnamefont {K.}~\bibnamefont
  {Watanabe}}, \bibinfo {author} {\bibfnamefont {T.}~\bibnamefont {Taniguchi}},
  \ and\ \bibinfo {author} {\bibfnamefont {H.}~\bibnamefont {Kanda}},\ }\href
  {\doibase 10.1038/nmat1134} {\bibfield  {journal} {\bibinfo  {journal}
  {Nature Materials}\ }\textbf {\bibinfo {volume} {3}},\ \bibinfo {pages} {404}
  (\bibinfo {year} {2004})}\BibitemShut {NoStop}%
\bibitem [{\citenamefont {Dean}\ \emph {et~al.}(2010)\citenamefont {Dean},
  \citenamefont {Young}, \citenamefont {Meric}, \citenamefont {Lee},
  \citenamefont {Wang}, \citenamefont {Sorgenfrei}, \citenamefont {Watanabe},
  \citenamefont {Taniguchi}, \citenamefont {Kim}, \citenamefont {Shepard},\
  and\ \citenamefont {Hone}}]{Dean2010BoronElectronics}%
  \BibitemOpen
  \bibfield  {author} {\bibinfo {author} {\bibfnamefont {C.~R.}\ \bibnamefont
  {Dean}}, \bibinfo {author} {\bibfnamefont {A.~F.}\ \bibnamefont {Young}},
  \bibinfo {author} {\bibfnamefont {I.}~\bibnamefont {Meric}}, \bibinfo
  {author} {\bibfnamefont {C.}~\bibnamefont {Lee}}, \bibinfo {author}
  {\bibfnamefont {L.}~\bibnamefont {Wang}}, \bibinfo {author} {\bibfnamefont
  {S.}~\bibnamefont {Sorgenfrei}}, \bibinfo {author} {\bibfnamefont
  {K.}~\bibnamefont {Watanabe}}, \bibinfo {author} {\bibfnamefont
  {T.}~\bibnamefont {Taniguchi}}, \bibinfo {author} {\bibfnamefont
  {P.}~\bibnamefont {Kim}}, \bibinfo {author} {\bibfnamefont {K.~L.}\
  \bibnamefont {Shepard}}, \ and\ \bibinfo {author} {\bibfnamefont
  {J.}~\bibnamefont {Hone}},\ }\href {\doibase 10.1038/nnano.2010.172}
  {\bibfield  {journal} {\bibinfo  {journal} {Nature Nanotechnology}\ }\textbf
  {\bibinfo {volume} {5}},\ \bibinfo {pages} {722} (\bibinfo {year}
  {2010})}\BibitemShut {NoStop}%
\bibitem [{\citenamefont {Yankowitz}\ \emph {et~al.}(2019)\citenamefont
  {Yankowitz}, \citenamefont {Ma}, \citenamefont {Jarillo-Herrero},\ and\
  \citenamefont {LeRoy}}]{Yankowitz2019VanNitride}%
  \BibitemOpen
  \bibfield  {author} {\bibinfo {author} {\bibfnamefont {M.}~\bibnamefont
  {Yankowitz}}, \bibinfo {author} {\bibfnamefont {Q.}~\bibnamefont {Ma}},
  \bibinfo {author} {\bibfnamefont {P.}~\bibnamefont {Jarillo-Herrero}}, \ and\
  \bibinfo {author} {\bibfnamefont {B.~J.}\ \bibnamefont {LeRoy}},\ }\bibfield
  {booktitle} {\emph {\bibinfo {booktitle} {Nature Reviews Physics}},\ }\href
  {\doibase 10.1038/s42254-018-0016-0} {\bibfield  {journal} {\bibinfo
  {journal} {Nature Reviews Physics}\ }\textbf {\bibinfo {volume} {1}},\
  \bibinfo {pages} {112} (\bibinfo {year} {2019})}\BibitemShut {NoStop}%
\bibitem [{\citenamefont {Cao}\ \emph {et~al.}(2018)\citenamefont {Cao},
  \citenamefont {Fatemi}, \citenamefont {Demir}, \citenamefont {Fang},
  \citenamefont {Tomarken}, \citenamefont {Luo}, \citenamefont
  {Sanchez-Yamagishi}, \citenamefont {Watanabe}, \citenamefont {Taniguchi},
  \citenamefont {Kaxiras}, \citenamefont {Ashoori},\ and\ \citenamefont
  {Jarillo-Herrero}}]{Cao2018CorrelatedSuperlattices}%
  \BibitemOpen
  \bibfield  {author} {\bibinfo {author} {\bibfnamefont {Y.}~\bibnamefont
  {Cao}}, \bibinfo {author} {\bibfnamefont {V.}~\bibnamefont {Fatemi}},
  \bibinfo {author} {\bibfnamefont {A.}~\bibnamefont {Demir}}, \bibinfo
  {author} {\bibfnamefont {S.}~\bibnamefont {Fang}}, \bibinfo {author}
  {\bibfnamefont {S.~L.}\ \bibnamefont {Tomarken}}, \bibinfo {author}
  {\bibfnamefont {J.~Y.}\ \bibnamefont {Luo}}, \bibinfo {author} {\bibfnamefont
  {J.~D.}\ \bibnamefont {Sanchez-Yamagishi}}, \bibinfo {author} {\bibfnamefont
  {K.}~\bibnamefont {Watanabe}}, \bibinfo {author} {\bibfnamefont
  {T.}~\bibnamefont {Taniguchi}}, \bibinfo {author} {\bibfnamefont
  {E.}~\bibnamefont {Kaxiras}}, \bibinfo {author} {\bibfnamefont {R.~C.}\
  \bibnamefont {Ashoori}}, \ and\ \bibinfo {author} {\bibfnamefont
  {P.}~\bibnamefont {Jarillo-Herrero}},\ }\href {\doibase 10.1038/nature26154}
  {\bibfield  {journal} {\bibinfo  {journal} {Nature}\ }\textbf {\bibinfo
  {volume} {556}},\ \bibinfo {pages} {80} (\bibinfo {year} {2018})}\BibitemShut
  {NoStop}%
\bibitem [{\citenamefont {Uri}\ \emph {et~al.}(2020)\citenamefont {Uri},
  \citenamefont {Grover}, \citenamefont {Cao}, \citenamefont {Crosse},
  \citenamefont {Bagani}, \citenamefont {Rodan-Legrain}, \citenamefont
  {Myasoedov}, \citenamefont {Watanabe}, \citenamefont {Taniguchi},
  \citenamefont {Moon}, \citenamefont {Koshino}, \citenamefont
  {Jarillo-Herrero},\ and\ \citenamefont {Zeldov}}]{Uri2020MappingGraphene}%
  \BibitemOpen
  \bibfield  {author} {\bibinfo {author} {\bibfnamefont {A.}~\bibnamefont
  {Uri}}, \bibinfo {author} {\bibfnamefont {S.}~\bibnamefont {Grover}},
  \bibinfo {author} {\bibfnamefont {Y.}~\bibnamefont {Cao}}, \bibinfo {author}
  {\bibfnamefont {J.~A.}\ \bibnamefont {Crosse}}, \bibinfo {author}
  {\bibfnamefont {K.}~\bibnamefont {Bagani}}, \bibinfo {author} {\bibfnamefont
  {D.}~\bibnamefont {Rodan-Legrain}}, \bibinfo {author} {\bibfnamefont
  {Y.}~\bibnamefont {Myasoedov}}, \bibinfo {author} {\bibfnamefont
  {K.}~\bibnamefont {Watanabe}}, \bibinfo {author} {\bibfnamefont
  {T.}~\bibnamefont {Taniguchi}}, \bibinfo {author} {\bibfnamefont
  {P.}~\bibnamefont {Moon}}, \bibinfo {author} {\bibfnamefont {M.}~\bibnamefont
  {Koshino}}, \bibinfo {author} {\bibfnamefont {P.}~\bibnamefont
  {Jarillo-Herrero}}, \ and\ \bibinfo {author} {\bibfnamefont {E.}~\bibnamefont
  {Zeldov}},\ }\href {\doibase 10.1038/s41586-020-2255-3} {\bibfield  {journal}
  {\bibinfo  {journal} {Nature}\ }\textbf {\bibinfo {volume} {581}},\ \bibinfo
  {pages} {47} (\bibinfo {year} {2020})}\BibitemShut {NoStop}%
\bibitem [{\citenamefont {Ye}\ \emph {et~al.}(2021)\citenamefont {Ye},
  \citenamefont {Islam}, \citenamefont {Zhang},\ and\ \citenamefont
  {Feng}}]{Ye2021UltrawideResonators}%
  \BibitemOpen
  \bibfield  {author} {\bibinfo {author} {\bibfnamefont {F.}~\bibnamefont
  {Ye}}, \bibinfo {author} {\bibfnamefont {A.}~\bibnamefont {Islam}}, \bibinfo
  {author} {\bibfnamefont {T.}~\bibnamefont {Zhang}}, \ and\ \bibinfo {author}
  {\bibfnamefont {P.~X.-L.}\ \bibnamefont {Feng}},\ }\href {\doibase
  10.1021/acs.nanolett.1c00610} {\bibfield  {journal} {\bibinfo  {journal}
  {Nano Letters}\ }\textbf {\bibinfo {volume} {21}},\ \bibinfo {pages} {5508}
  (\bibinfo {year} {2021})}\BibitemShut {NoStop}%
\bibitem [{\citenamefont {Kim}\ \emph {et~al.}(2020)\citenamefont {Kim},
  \citenamefont {Annevelink}, \citenamefont {Han}, \citenamefont {Yu},
  \citenamefont {Huang}, \citenamefont {Ertekin},\ and\ \citenamefont {van~der
  Zande}}]{Kim2020StochasticInterfaces}%
  \BibitemOpen
  \bibfield  {author} {\bibinfo {author} {\bibfnamefont {S.}~\bibnamefont
  {Kim}}, \bibinfo {author} {\bibfnamefont {E.}~\bibnamefont {Annevelink}},
  \bibinfo {author} {\bibfnamefont {E.}~\bibnamefont {Han}}, \bibinfo {author}
  {\bibfnamefont {J.}~\bibnamefont {Yu}}, \bibinfo {author} {\bibfnamefont
  {P.~Y.}\ \bibnamefont {Huang}}, \bibinfo {author} {\bibfnamefont
  {E.}~\bibnamefont {Ertekin}}, \ and\ \bibinfo {author} {\bibfnamefont
  {A.~M.}\ \bibnamefont {van~der Zande}},\ }\href {\doibase
  10.1021/acs.nanolett.9b04619} {\bibfield  {journal} {\bibinfo  {journal}
  {Nano Letters}\ }\textbf {\bibinfo {volume} {20}},\ \bibinfo {pages} {1201}
  (\bibinfo {year} {2020})}\BibitemShut {NoStop}%
\bibitem [{\citenamefont {Kim}\ \emph {et~al.}(2018{\natexlab{a}})\citenamefont
  {Kim}, \citenamefont {Yu},\ and\ \citenamefont {van~der
  Zande}}]{Kim2018Nano-electromechanicalBimorphs}%
  \BibitemOpen
  \bibfield  {author} {\bibinfo {author} {\bibfnamefont {S.}~\bibnamefont
  {Kim}}, \bibinfo {author} {\bibfnamefont {J.}~\bibnamefont {Yu}}, \ and\
  \bibinfo {author} {\bibfnamefont {A.~M.}\ \bibnamefont {van~der Zande}},\
  }\href {\doibase 10.1021/acs.nanolett.8b01926} {\bibfield  {journal}
  {\bibinfo  {journal} {Nano Letters}\ }\textbf {\bibinfo {volume} {18}},\
  \bibinfo {pages} {6686} (\bibinfo {year} {2018}{\natexlab{a}})}\BibitemShut
  {NoStop}%
\bibitem [{\citenamefont {Kumar}\ \emph {et~al.}(2020)\citenamefont {Kumar},
  \citenamefont {Session}, \citenamefont {Tsuchikawa}, \citenamefont {Homer},
  \citenamefont {Paas}, \citenamefont {Watanabe}, \citenamefont {Taniguchi},\
  and\ \citenamefont {Deshpande}}]{Kumar2020CircularHeterostructures}%
  \BibitemOpen
  \bibfield  {author} {\bibinfo {author} {\bibfnamefont {R.}~\bibnamefont
  {Kumar}}, \bibinfo {author} {\bibfnamefont {D.~W.}\ \bibnamefont {Session}},
  \bibinfo {author} {\bibfnamefont {R.}~\bibnamefont {Tsuchikawa}}, \bibinfo
  {author} {\bibfnamefont {M.}~\bibnamefont {Homer}}, \bibinfo {author}
  {\bibfnamefont {H.}~\bibnamefont {Paas}}, \bibinfo {author} {\bibfnamefont
  {K.}~\bibnamefont {Watanabe}}, \bibinfo {author} {\bibfnamefont
  {T.}~\bibnamefont {Taniguchi}}, \ and\ \bibinfo {author} {\bibfnamefont
  {V.~V.}\ \bibnamefont {Deshpande}},\ }\href {\doibase 10.1063/5.0024583}
  {\bibfield  {journal} {\bibinfo  {journal} {Applied Physics Letters}\
  }\textbf {\bibinfo {volume} {117}},\ \bibinfo {pages} {183103} (\bibinfo
  {year} {2020})}\BibitemShut {NoStop}%
\bibitem [{\citenamefont {Yu}\ \emph {et~al.}(2021{\natexlab{b}})\citenamefont
  {Yu}, \citenamefont {Han}, \citenamefont {Hossain}, \citenamefont {Watanabe},
  \citenamefont {Taniguchi}, \citenamefont {Ertekin}, \citenamefont {van~der
  Zande},\ and\ \citenamefont {Huang}}]{Yu2021DesigningHeterostructures}%
  \BibitemOpen
  \bibfield  {author} {\bibinfo {author} {\bibfnamefont {J.}~\bibnamefont
  {Yu}}, \bibinfo {author} {\bibfnamefont {E.}~\bibnamefont {Han}}, \bibinfo
  {author} {\bibfnamefont {M.~A.}\ \bibnamefont {Hossain}}, \bibinfo {author}
  {\bibfnamefont {K.}~\bibnamefont {Watanabe}}, \bibinfo {author}
  {\bibfnamefont {T.}~\bibnamefont {Taniguchi}}, \bibinfo {author}
  {\bibfnamefont {E.}~\bibnamefont {Ertekin}}, \bibinfo {author} {\bibfnamefont
  {A.~M.}\ \bibnamefont {van~der Zande}}, \ and\ \bibinfo {author}
  {\bibfnamefont {P.~Y.}\ \bibnamefont {Huang}},\ }\href {\doibase
  https://doi.org/10.1002/adma.202007269} {\bibfield  {journal} {\bibinfo
  {journal} {Advanced Materials}\ }\textbf {\bibinfo {volume} {33}},\ \bibinfo
  {pages} {2007269} (\bibinfo {year} {2021}{\natexlab{b}})}\BibitemShut
  {NoStop}%
\bibitem [{\citenamefont {Han}\ \emph {et~al.}(2020)\citenamefont {Han},
  \citenamefont {Yu}, \citenamefont {Annevelink}, \citenamefont {Son},
  \citenamefont {Kang}, \citenamefont {Watanabe}, \citenamefont {Taniguchi},
  \citenamefont {Ertekin}, \citenamefont {Huang},\ and\ \citenamefont {van~der
  Zande}}]{han2020ultrasoft}%
  \BibitemOpen
  \bibfield  {author} {\bibinfo {author} {\bibfnamefont {E.}~\bibnamefont
  {Han}}, \bibinfo {author} {\bibfnamefont {J.}~\bibnamefont {Yu}}, \bibinfo
  {author} {\bibfnamefont {E.}~\bibnamefont {Annevelink}}, \bibinfo {author}
  {\bibfnamefont {J.}~\bibnamefont {Son}}, \bibinfo {author} {\bibfnamefont
  {D.~A.}\ \bibnamefont {Kang}}, \bibinfo {author} {\bibfnamefont
  {K.}~\bibnamefont {Watanabe}}, \bibinfo {author} {\bibfnamefont
  {T.}~\bibnamefont {Taniguchi}}, \bibinfo {author} {\bibfnamefont
  {E.}~\bibnamefont {Ertekin}}, \bibinfo {author} {\bibfnamefont {P.~Y.}\
  \bibnamefont {Huang}}, \ and\ \bibinfo {author} {\bibfnamefont {A.~M.}\
  \bibnamefont {van~der Zande}},\ }\href@noop {} {\bibfield  {journal}
  {\bibinfo  {journal} {Nature materials}\ }\textbf {\bibinfo {volume} {19}},\
  \bibinfo {pages} {305} (\bibinfo {year} {2020})}\BibitemShut {NoStop}%
\bibitem [{\citenamefont {Frank}\ \emph {et~al.}(2007)\citenamefont {Frank},
  \citenamefont {Tanenbaum}, \citenamefont {Zande},\ and\ \citenamefont
  {McEuen}}]{Frank2007MechanicalSheets}%
  \BibitemOpen
  \bibfield  {author} {\bibinfo {author} {\bibfnamefont {I.~W.}\ \bibnamefont
  {Frank}}, \bibinfo {author} {\bibfnamefont {D.~M.}\ \bibnamefont
  {Tanenbaum}}, \bibinfo {author} {\bibfnamefont {A.~M. v.~d.}\ \bibnamefont
  {Zande}}, \ and\ \bibinfo {author} {\bibfnamefont {P.~L.}\ \bibnamefont
  {McEuen}},\ }\href {\doibase 10.1116/1.2789446} {\bibfield  {journal}
  {\bibinfo  {journal} {Journal of Vacuum Science {\&} Technology B:
  Microelectronics and Nanometer Structures Processing, Measurement, and
  Phenomena}\ }\textbf {\bibinfo {volume} {25}},\ \bibinfo {pages} {2558}
  (\bibinfo {year} {2007})}\BibitemShut {NoStop}%
\bibitem [{\citenamefont {Cao}\ \emph {et~al.}(2014)\citenamefont {Cao},
  \citenamefont {Wang}, \citenamefont {Gao}, \citenamefont {Tao}, \citenamefont
  {Suk}, \citenamefont {Ruoff}, \citenamefont {Akinwande}, \citenamefont
  {Huang},\ and\ \citenamefont {Liechti}}]{Cao2014AGraphene}%
  \BibitemOpen
  \bibfield  {author} {\bibinfo {author} {\bibfnamefont {Z.}~\bibnamefont
  {Cao}}, \bibinfo {author} {\bibfnamefont {P.}~\bibnamefont {Wang}}, \bibinfo
  {author} {\bibfnamefont {W.}~\bibnamefont {Gao}}, \bibinfo {author}
  {\bibfnamefont {L.}~\bibnamefont {Tao}}, \bibinfo {author} {\bibfnamefont
  {J.~W.}\ \bibnamefont {Suk}}, \bibinfo {author} {\bibfnamefont {R.~S.}\
  \bibnamefont {Ruoff}}, \bibinfo {author} {\bibfnamefont {D.}~\bibnamefont
  {Akinwande}}, \bibinfo {author} {\bibfnamefont {R.}~\bibnamefont {Huang}}, \
  and\ \bibinfo {author} {\bibfnamefont {K.~M.}\ \bibnamefont {Liechti}},\
  }\href {\doibase 10.1016/J.CARBON.2013.12.041} {\bibfield  {journal}
  {\bibinfo  {journal} {Carbon}\ }\textbf {\bibinfo {volume} {69}},\ \bibinfo
  {pages} {390} (\bibinfo {year} {2014})}\BibitemShut {NoStop}%
\bibitem [{\citenamefont {Wang}\ \emph
  {et~al.}(2019{\natexlab{a}})\citenamefont {Wang}, \citenamefont {Dai},
  \citenamefont {Xiao}, \citenamefont {Feng}, \citenamefont {Weng},
  \citenamefont {Liu}, \citenamefont {Xu}, \citenamefont {Huang},\ and\
  \citenamefont {Zhang}}]{Wang2019BendingMaterials}%
  \BibitemOpen
  \bibfield  {author} {\bibinfo {author} {\bibfnamefont {G.}~\bibnamefont
  {Wang}}, \bibinfo {author} {\bibfnamefont {Z.}~\bibnamefont {Dai}}, \bibinfo
  {author} {\bibfnamefont {J.}~\bibnamefont {Xiao}}, \bibinfo {author}
  {\bibfnamefont {S.}~\bibnamefont {Feng}}, \bibinfo {author} {\bibfnamefont
  {C.}~\bibnamefont {Weng}}, \bibinfo {author} {\bibfnamefont {L.}~\bibnamefont
  {Liu}}, \bibinfo {author} {\bibfnamefont {Z.}~\bibnamefont {Xu}}, \bibinfo
  {author} {\bibfnamefont {R.}~\bibnamefont {Huang}}, \ and\ \bibinfo {author}
  {\bibfnamefont {Z.}~\bibnamefont {Zhang}},\ }\href {\doibase
  10.1103/PhysRevLett.123.116101} {\bibfield  {journal} {\bibinfo  {journal}
  {Phys. Rev. Lett.}\ }\textbf {\bibinfo {volume} {123}},\ \bibinfo {pages}
  {116101} (\bibinfo {year} {2019}{\natexlab{a}})}\BibitemShut {NoStop}%
\bibitem [{\citenamefont {Zong}\ \emph {et~al.}(2010)\citenamefont {Zong},
  \citenamefont {Chen}, \citenamefont {Dokmeci},\ and\ \citenamefont
  {Wan}}]{Zong2010DirectNanoparticles}%
  \BibitemOpen
  \bibfield  {author} {\bibinfo {author} {\bibfnamefont {Z.}~\bibnamefont
  {Zong}}, \bibinfo {author} {\bibfnamefont {C.-L.}\ \bibnamefont {Chen}},
  \bibinfo {author} {\bibfnamefont {M.~R.}\ \bibnamefont {Dokmeci}}, \ and\
  \bibinfo {author} {\bibfnamefont {K.-T.}\ \bibnamefont {Wan}},\ }\href
  {\doibase 10.1063/1.3294960} {\bibfield  {journal} {\bibinfo  {journal} {J.
  Appl. Phys}\ }\textbf {\bibinfo {volume} {107}},\ \bibinfo {pages} {026104}
  (\bibinfo {year} {2010})}\BibitemShut {NoStop}%
\bibitem [{\citenamefont {Yoon}\ \emph {et~al.}(2012)\citenamefont {Yoon},
  \citenamefont {Shin}, \citenamefont {Kim}, \citenamefont {Mun}, \citenamefont
  {Kim},\ and\ \citenamefont {Cho}}]{Yoon2012DirectProcess}%
  \BibitemOpen
  \bibfield  {author} {\bibinfo {author} {\bibfnamefont {T.}~\bibnamefont
  {Yoon}}, \bibinfo {author} {\bibfnamefont {W.~C.}\ \bibnamefont {Shin}},
  \bibinfo {author} {\bibfnamefont {T.~Y.}\ \bibnamefont {Kim}}, \bibinfo
  {author} {\bibfnamefont {J.~H.}\ \bibnamefont {Mun}}, \bibinfo {author}
  {\bibfnamefont {T.-S.}\ \bibnamefont {Kim}}, \ and\ \bibinfo {author}
  {\bibfnamefont {B.~J.}\ \bibnamefont {Cho}},\ }\href {\doibase
  10.1021/NL204123H} {\bibfield  {journal} {\bibinfo  {journal} {Nano Letters}\
  }\textbf {\bibinfo {volume} {12}},\ \bibinfo {pages} {1448} (\bibinfo {year}
  {2012})}\BibitemShut {NoStop}%
\bibitem [{\citenamefont {Dai}\ \emph {et~al.}(2020{\natexlab{a}})\citenamefont
  {Dai}, \citenamefont {Lu}, \citenamefont {Liechti},\ and\ \citenamefont
  {Huang}}]{dai2020mechanicsopportunities}%
  \BibitemOpen
  \bibfield  {author} {\bibinfo {author} {\bibfnamefont {Z.}~\bibnamefont
  {Dai}}, \bibinfo {author} {\bibfnamefont {N.}~\bibnamefont {Lu}}, \bibinfo
  {author} {\bibfnamefont {K.~M.}\ \bibnamefont {Liechti}}, \ and\ \bibinfo
  {author} {\bibfnamefont {R.}~\bibnamefont {Huang}},\ }\href {\doibase
  10.1016/j.cossms.2020.100837} {\bibfield  {journal} {\bibinfo  {journal}
  {Current Opinion in Solid State and Materials Science}\ }\textbf {\bibinfo
  {volume} {24}},\ \bibinfo {pages} {100837} (\bibinfo {year}
  {2020}{\natexlab{a}})}\BibitemShut {NoStop}%
\bibitem [{\citenamefont {Pereira}\ and\ \citenamefont
  {Neto}(2009)}]{Pereira2009StrainStructure}%
  \BibitemOpen
  \bibfield  {author} {\bibinfo {author} {\bibfnamefont {V.~M.}\ \bibnamefont
  {Pereira}}\ and\ \bibinfo {author} {\bibfnamefont {A.~H.~C.}\ \bibnamefont
  {Neto}},\ }\href {\doibase 10.1103/PhysRevLett.103.046801} {\bibfield
  {journal} {\bibinfo  {journal} {Physical Review Letters}\ }\textbf {\bibinfo
  {volume} {103}},\ \bibinfo {pages} {046801} (\bibinfo {year}
  {2009})}\BibitemShut {NoStop}%
\bibitem [{\citenamefont {Ferrari}\ \emph {et~al.}(2021)\citenamefont
  {Ferrari}, \citenamefont {Kim},\ and\ \citenamefont {van~der
  Zande}}]{Ferrari2021DissipationResonators}%
  \BibitemOpen
  \bibfield  {author} {\bibinfo {author} {\bibfnamefont {P.~F.}\ \bibnamefont
  {Ferrari}}, \bibinfo {author} {\bibfnamefont {S.}~\bibnamefont {Kim}}, \ and\
  \bibinfo {author} {\bibfnamefont {A.~M.}\ \bibnamefont {van~der Zande}},\
  }\href {\doibase 10.1021/acs.nanolett.1c02369} {\bibfield  {journal}
  {\bibinfo  {journal} {Nano Letters}\ }\textbf {\bibinfo {volume} {21}},\
  \bibinfo {pages} {8058} (\bibinfo {year} {2021})}\BibitemShut {NoStop}%
\bibitem [{\citenamefont {Will}\ \emph {et~al.}(2017)\citenamefont {Will},
  \citenamefont {Hamer}, \citenamefont {M{\"{u}}ller}, \citenamefont {Noury},
  \citenamefont {Weber}, \citenamefont {Bachtold}, \citenamefont {Gorbachev},
  \citenamefont {Stampfer},\ and\ \citenamefont
  {G{\"{u}}ttinger}}]{Will2017HighResonator}%
  \BibitemOpen
  \bibfield  {author} {\bibinfo {author} {\bibfnamefont {M.}~\bibnamefont
  {Will}}, \bibinfo {author} {\bibfnamefont {M.}~\bibnamefont {Hamer}},
  \bibinfo {author} {\bibfnamefont {M.}~\bibnamefont {M{\"{u}}ller}}, \bibinfo
  {author} {\bibfnamefont {A.}~\bibnamefont {Noury}}, \bibinfo {author}
  {\bibfnamefont {P.}~\bibnamefont {Weber}}, \bibinfo {author} {\bibfnamefont
  {A.}~\bibnamefont {Bachtold}}, \bibinfo {author} {\bibfnamefont {R.~V.}\
  \bibnamefont {Gorbachev}}, \bibinfo {author} {\bibfnamefont {C.}~\bibnamefont
  {Stampfer}}, \ and\ \bibinfo {author} {\bibfnamefont {J.}~\bibnamefont
  {G{\"{u}}ttinger}},\ }\href {\doibase 10.1021/acs.nanolett.7b01845}
  {\bibfield  {journal} {\bibinfo  {journal} {Nano Letters}\ }\textbf {\bibinfo
  {volume} {17}},\ \bibinfo {pages} {5950} (\bibinfo {year}
  {2017})}\BibitemShut {NoStop}%
\bibitem [{\citenamefont {Jia}\ \emph {et~al.}(2019)\citenamefont {Jia},
  \citenamefont {Chen}, \citenamefont {Qiao}, \citenamefont {Zhang},
  \citenamefont {Zheng}, \citenamefont {Xue}, \citenamefont {Liang},
  \citenamefont {Tian}, \citenamefont {He}, \citenamefont {Di} \emph
  {et~al.}}]{jia2019programmable}%
  \BibitemOpen
  \bibfield  {author} {\bibinfo {author} {\bibfnamefont {P.}~\bibnamefont
  {Jia}}, \bibinfo {author} {\bibfnamefont {W.}~\bibnamefont {Chen}}, \bibinfo
  {author} {\bibfnamefont {J.}~\bibnamefont {Qiao}}, \bibinfo {author}
  {\bibfnamefont {M.}~\bibnamefont {Zhang}}, \bibinfo {author} {\bibfnamefont
  {X.}~\bibnamefont {Zheng}}, \bibinfo {author} {\bibfnamefont
  {Z.}~\bibnamefont {Xue}}, \bibinfo {author} {\bibfnamefont {R.}~\bibnamefont
  {Liang}}, \bibinfo {author} {\bibfnamefont {C.}~\bibnamefont {Tian}},
  \bibinfo {author} {\bibfnamefont {L.}~\bibnamefont {He}}, \bibinfo {author}
  {\bibfnamefont {Z.}~\bibnamefont {Di}},  \emph {et~al.},\ }\href@noop {}
  {\bibfield  {journal} {\bibinfo  {journal} {Nature communications}\ }\textbf
  {\bibinfo {volume} {10}},\ \bibinfo {pages} {3127} (\bibinfo {year}
  {2019})}\BibitemShut {NoStop}%
\bibitem [{\citenamefont {De~Juan}\ \emph {et~al.}(2011)\citenamefont
  {De~Juan}, \citenamefont {Cortijo}, \citenamefont {Vozmediano},\ and\
  \citenamefont {Cano}}]{de2011aharonov}%
  \BibitemOpen
  \bibfield  {author} {\bibinfo {author} {\bibfnamefont {F.}~\bibnamefont
  {De~Juan}}, \bibinfo {author} {\bibfnamefont {A.}~\bibnamefont {Cortijo}},
  \bibinfo {author} {\bibfnamefont {M.~A.}\ \bibnamefont {Vozmediano}}, \ and\
  \bibinfo {author} {\bibfnamefont {A.}~\bibnamefont {Cano}},\ }\href@noop {}
  {\bibfield  {journal} {\bibinfo  {journal} {Nature Physics}\ }\textbf
  {\bibinfo {volume} {7}},\ \bibinfo {pages} {810} (\bibinfo {year}
  {2011})}\BibitemShut {NoStop}%
\bibitem [{\citenamefont {Griffith}\ and\ \citenamefont
  {Taylor}(1921)}]{griffith_vi_1921}%
  \BibitemOpen
  \bibfield  {author} {\bibinfo {author} {\bibfnamefont {A.~A.}\ \bibnamefont
  {Griffith}}\ and\ \bibinfo {author} {\bibfnamefont {G.~I.}\ \bibnamefont
  {Taylor}},\ }\href {\doibase 10.1098/rsta.1921.0006} {\bibfield  {journal}
  {\bibinfo  {journal} {Philosophical Transactions of the Royal Society of
  London. Series A, Containing Papers of a Mathematical or Physical Character}\
  }\textbf {\bibinfo {volume} {221}},\ \bibinfo {pages} {163} (\bibinfo {year}
  {1921})}\BibitemShut {NoStop}%
\bibitem [{\citenamefont {Khestanova}\ \emph {et~al.}(2016)\citenamefont
  {Khestanova}, \citenamefont {Guinea}, \citenamefont {Fumagalli},
  \citenamefont {Geim},\ and\ \citenamefont
  {Grigorieva}}]{Khestanova2016UniversalHeterostructures}%
  \BibitemOpen
  \bibfield  {author} {\bibinfo {author} {\bibfnamefont {E.}~\bibnamefont
  {Khestanova}}, \bibinfo {author} {\bibfnamefont {F.}~\bibnamefont {Guinea}},
  \bibinfo {author} {\bibfnamefont {L.}~\bibnamefont {Fumagalli}}, \bibinfo
  {author} {\bibfnamefont {A.~K.}\ \bibnamefont {Geim}}, \ and\ \bibinfo
  {author} {\bibfnamefont {I.~V.}\ \bibnamefont {Grigorieva}},\ }\href
  {\doibase 10.1038/ncomms12587} {\bibfield  {journal} {\bibinfo  {journal}
  {Nature Communications}\ }\textbf {\bibinfo {volume} {7}},\ \bibinfo {pages}
  {12587} (\bibinfo {year} {2016})}\BibitemShut {NoStop}%
\bibitem [{\citenamefont {Singh}\ \emph {et~al.}(2010)\citenamefont {Singh},
  \citenamefont {Sengupta}, \citenamefont {Solanki}, \citenamefont {Dhall},
  \citenamefont {Allain}, \citenamefont {Dhara}, \citenamefont {Pant},\ and\
  \citenamefont {Deshmukh}}]{Singh2010ProbingResonators}%
  \BibitemOpen
  \bibfield  {author} {\bibinfo {author} {\bibfnamefont {V.}~\bibnamefont
  {Singh}}, \bibinfo {author} {\bibfnamefont {S.}~\bibnamefont {Sengupta}},
  \bibinfo {author} {\bibfnamefont {H.~S.}\ \bibnamefont {Solanki}}, \bibinfo
  {author} {\bibfnamefont {R.}~\bibnamefont {Dhall}}, \bibinfo {author}
  {\bibfnamefont {A.}~\bibnamefont {Allain}}, \bibinfo {author} {\bibfnamefont
  {S.}~\bibnamefont {Dhara}}, \bibinfo {author} {\bibfnamefont
  {P.}~\bibnamefont {Pant}}, \ and\ \bibinfo {author} {\bibfnamefont {M.~M.}\
  \bibnamefont {Deshmukh}},\ }\href {\doibase 10.1088/0957-4484/21/16/165204}
  {\bibfield  {journal} {\bibinfo  {journal} {Nanotechnology}\ }\textbf
  {\bibinfo {volume} {21}},\ \bibinfo {pages} {165204} (\bibinfo {year}
  {2010})}\BibitemShut {NoStop}%
\bibitem [{\citenamefont {Van Der~Zande}\ \emph {et~al.}(2010)\citenamefont
  {Van Der~Zande}, \citenamefont {Barton}, \citenamefont {Alden}, \citenamefont
  {Ruiz-Vargas}, \citenamefont {Whitney}, \citenamefont {Pham}, \citenamefont
  {Park}, \citenamefont {Parpia}, \citenamefont {Craighead},\ and\
  \citenamefont {McEuen}}]{VanDerZande2010Large-scaleResonators}%
  \BibitemOpen
  \bibfield  {author} {\bibinfo {author} {\bibfnamefont {A.~M.}\ \bibnamefont
  {Van Der~Zande}}, \bibinfo {author} {\bibfnamefont {R.~A.}\ \bibnamefont
  {Barton}}, \bibinfo {author} {\bibfnamefont {J.~S.}\ \bibnamefont {Alden}},
  \bibinfo {author} {\bibfnamefont {C.~S.}\ \bibnamefont {Ruiz-Vargas}},
  \bibinfo {author} {\bibfnamefont {W.~S.}\ \bibnamefont {Whitney}}, \bibinfo
  {author} {\bibfnamefont {P.~H.}\ \bibnamefont {Pham}}, \bibinfo {author}
  {\bibfnamefont {J.}~\bibnamefont {Park}}, \bibinfo {author} {\bibfnamefont
  {J.~M.}\ \bibnamefont {Parpia}}, \bibinfo {author} {\bibfnamefont {H.~G.}\
  \bibnamefont {Craighead}}, \ and\ \bibinfo {author} {\bibfnamefont {P.~L.}\
  \bibnamefont {McEuen}},\ }\href {\doibase 10.1021/nl102713c} {\bibfield
  {journal} {\bibinfo  {journal} {Nano Letters}\ }\textbf {\bibinfo {volume}
  {10}},\ \bibinfo {pages} {4869} (\bibinfo {year} {2010})}\BibitemShut
  {NoStop}%
\bibitem [{\citenamefont {Chen}\ \emph {et~al.}(2009)\citenamefont {Chen},
  \citenamefont {Rosenblatt}, \citenamefont {Bolotin}, \citenamefont {Kalb},
  \citenamefont {Kim}, \citenamefont {Kymissis}, \citenamefont {Stormer},
  \citenamefont {Heinz},\ and\ \citenamefont
  {Hone}}]{Chen2009PerformanceReadout}%
  \BibitemOpen
  \bibfield  {author} {\bibinfo {author} {\bibfnamefont {C.}~\bibnamefont
  {Chen}}, \bibinfo {author} {\bibfnamefont {S.}~\bibnamefont {Rosenblatt}},
  \bibinfo {author} {\bibfnamefont {K.~I.}\ \bibnamefont {Bolotin}}, \bibinfo
  {author} {\bibfnamefont {W.}~\bibnamefont {Kalb}}, \bibinfo {author}
  {\bibfnamefont {P.}~\bibnamefont {Kim}}, \bibinfo {author} {\bibfnamefont
  {I.}~\bibnamefont {Kymissis}}, \bibinfo {author} {\bibfnamefont {H.~L.}\
  \bibnamefont {Stormer}}, \bibinfo {author} {\bibfnamefont {T.~F.}\
  \bibnamefont {Heinz}}, \ and\ \bibinfo {author} {\bibfnamefont
  {J.}~\bibnamefont {Hone}},\ }\href {\doibase 10.1038/nnano.2009.267}
  {\bibfield  {journal} {\bibinfo  {journal} {Nature Nanotechnology}\ }\textbf
  {\bibinfo {volume} {4}},\ \bibinfo {pages} {861} (\bibinfo {year}
  {2009})}\BibitemShut {NoStop}%
\bibitem [{\citenamefont {G{\"{u}}ttinger}\ \emph {et~al.}(2017)\citenamefont
  {G{\"{u}}ttinger}, \citenamefont {Noury}, \citenamefont {Weber},
  \citenamefont {Eriksson}, \citenamefont {Lagoin}, \citenamefont {Moser},
  \citenamefont {Eichler}, \citenamefont {Wallraff}, \citenamefont {Isacsson},\
  and\ \citenamefont {Bachtold}}]{Guttinger2017Energy-dependentResonators}%
  \BibitemOpen
  \bibfield  {author} {\bibinfo {author} {\bibfnamefont {J.}~\bibnamefont
  {G{\"{u}}ttinger}}, \bibinfo {author} {\bibfnamefont {A.}~\bibnamefont
  {Noury}}, \bibinfo {author} {\bibfnamefont {P.}~\bibnamefont {Weber}},
  \bibinfo {author} {\bibfnamefont {A.~M.}\ \bibnamefont {Eriksson}}, \bibinfo
  {author} {\bibfnamefont {C.}~\bibnamefont {Lagoin}}, \bibinfo {author}
  {\bibfnamefont {J.}~\bibnamefont {Moser}}, \bibinfo {author} {\bibfnamefont
  {C.}~\bibnamefont {Eichler}}, \bibinfo {author} {\bibfnamefont
  {A.}~\bibnamefont {Wallraff}}, \bibinfo {author} {\bibfnamefont
  {A.}~\bibnamefont {Isacsson}}, \ and\ \bibinfo {author} {\bibfnamefont
  {A.}~\bibnamefont {Bachtold}},\ }\href {\doibase 10.1038/nnano.2017.86}
  {\bibfield  {journal} {\bibinfo  {journal} {Nature Nanotechnology}\ }\textbf
  {\bibinfo {volume} {12}},\ \bibinfo {pages} {631} (\bibinfo {year}
  {2017})}\BibitemShut {NoStop}%
\bibitem [{\citenamefont {Kozinsky}\ \emph {et~al.}(2006)\citenamefont
  {Kozinsky}, \citenamefont {Postma}, \citenamefont {Bargatin},\ and\
  \citenamefont {Roukes}}]{Kozinsky2006TuningResonators}%
  \BibitemOpen
  \bibfield  {author} {\bibinfo {author} {\bibfnamefont {I.}~\bibnamefont
  {Kozinsky}}, \bibinfo {author} {\bibfnamefont {H.~W.~C.}\ \bibnamefont
  {Postma}}, \bibinfo {author} {\bibfnamefont {I.}~\bibnamefont {Bargatin}}, \
  and\ \bibinfo {author} {\bibfnamefont {M.~L.}\ \bibnamefont {Roukes}},\
  }\href {\doibase 10.1063/1.2209211} {\bibfield  {journal} {\bibinfo
  {journal} {Applied Physics Letters}\ }\textbf {\bibinfo {volume} {88}},\
  \bibinfo {pages} {253101} (\bibinfo {year} {2006})}\BibitemShut {NoStop}%
\bibitem [{\citenamefont
  {Anderson}(2005)}]{TedL.Anderson2005FractureApplications}%
  \BibitemOpen
  \bibfield  {author} {\bibinfo {author} {\bibfnamefont {T.~L.}\ \bibnamefont
  {Anderson}},\ }\href {https://books.google.co.in/books?id=C5fMBQAAQBAJ}
  {\emph {\bibinfo {title} {Fracture Mechanics: Fundamentals and
  Applications}}},\ \bibinfo {edition} {3rd}\ ed.\ (\bibinfo  {publisher} {CRC
  Press},\ \bibinfo {year} {2005})\BibitemShut {NoStop}%
\bibitem [{\citenamefont {Jain}\ \emph {et~al.}(2018)\citenamefont {Jain},
  \citenamefont {Bharadwaj}, \citenamefont {Heeg}, \citenamefont {Parzefall},
  \citenamefont {Taniguchi}, \citenamefont {Watanabe},\ and\ \citenamefont
  {Novotny}}]{Jain2018MinimizingPDMS}%
  \BibitemOpen
  \bibfield  {author} {\bibinfo {author} {\bibfnamefont {A.}~\bibnamefont
  {Jain}}, \bibinfo {author} {\bibfnamefont {P.}~\bibnamefont {Bharadwaj}},
  \bibinfo {author} {\bibfnamefont {S.}~\bibnamefont {Heeg}}, \bibinfo {author}
  {\bibfnamefont {M.}~\bibnamefont {Parzefall}}, \bibinfo {author}
  {\bibfnamefont {T.}~\bibnamefont {Taniguchi}}, \bibinfo {author}
  {\bibfnamefont {K.}~\bibnamefont {Watanabe}}, \ and\ \bibinfo {author}
  {\bibfnamefont {L.}~\bibnamefont {Novotny}},\ }\href {\doibase
  10.1088/1361-6528/aabd90} {\bibfield  {journal} {\bibinfo  {journal}
  {Nanotechnology}\ }\textbf {\bibinfo {volume} {29}},\ \bibinfo {pages}
  {265203} (\bibinfo {year} {2018})}\BibitemShut {NoStop}%
\bibitem [{\citenamefont {Sanchez}\ \emph {et~al.}(2018)\citenamefont
  {Sanchez}, \citenamefont {Dai}, \citenamefont {Wang}, \citenamefont
  {Cantu-Chavez}, \citenamefont {Brennan}, \citenamefont {Huang},\ and\
  \citenamefont {Lu}}]{Sanchez2018MechanicsCrystals}%
  \BibitemOpen
  \bibfield  {author} {\bibinfo {author} {\bibfnamefont {D.~A.}\ \bibnamefont
  {Sanchez}}, \bibinfo {author} {\bibfnamefont {Z.}~\bibnamefont {Dai}},
  \bibinfo {author} {\bibfnamefont {P.}~\bibnamefont {Wang}}, \bibinfo {author}
  {\bibfnamefont {A.}~\bibnamefont {Cantu-Chavez}}, \bibinfo {author}
  {\bibfnamefont {C.~J.}\ \bibnamefont {Brennan}}, \bibinfo {author}
  {\bibfnamefont {R.}~\bibnamefont {Huang}}, \ and\ \bibinfo {author}
  {\bibfnamefont {N.}~\bibnamefont {Lu}},\ }\href {\doibase
  10.1073/pnas.1801551115} {\bibfield  {journal} {\bibinfo  {journal}
  {Proceedings of the National Academy of Sciences of the United States of
  America}\ }\textbf {\bibinfo {volume} {115}},\ \bibinfo {pages} {7884}
  (\bibinfo {year} {2018})}\BibitemShut {NoStop}%
\bibitem [{\citenamefont {Gvirtzman}\ and\ \citenamefont
  {Fineberg}(2021)}]{gvirtzman_nucleation_2021}%
  \BibitemOpen
  \bibfield  {author} {\bibinfo {author} {\bibfnamefont {S.}~\bibnamefont
  {Gvirtzman}}\ and\ \bibinfo {author} {\bibfnamefont {J.}~\bibnamefont
  {Fineberg}},\ }\href {\doibase 10.1038/s41567-021-01299-9} {\bibfield
  {journal} {\bibinfo  {journal} {Nature Physics}\ }\textbf {\bibinfo {volume}
  {17}},\ \bibinfo {pages} {1037} (\bibinfo {year} {2021})}\BibitemShut
  {NoStop}%
\bibitem [{\citenamefont
  {Malthe-Sørenssen}(2021)}]{malthe-sorenssen_onset_2021}%
  \BibitemOpen
  \bibfield  {author} {\bibinfo {author} {\bibfnamefont {A.}~\bibnamefont
  {Malthe-Sørenssen}},\ }\href {\doibase 10.1038/s41567-021-01312-1}
  {\bibfield  {journal} {\bibinfo  {journal} {Nature Physics}\ }\textbf
  {\bibinfo {volume} {17}},\ \bibinfo {pages} {983} (\bibinfo {year}
  {2021})}\BibitemShut {NoStop}%

\end{thebibliography}

\begin{thebibliography}{52}%
\makeatletter
\providecommand \@ifxundefined [1]{%
 \@ifx{#1\undefined}
}%
\providecommand \@ifnum [1]{%
 \ifnum #1\expandafter \@firstoftwo
 \else \expandafter \@secondoftwo
 \fi
}%
\providecommand \@ifx [1]{%
 \ifx #1\expandafter \@firstoftwo
 \else \expandafter \@secondoftwo
 \fi
}%
\providecommand \natexlab [1]{#1}%
\providecommand \enquote  [1]{``#1''}%
\providecommand \bibnamefont  [1]{#1}%
\providecommand \bibfnamefont [1]{#1}%
\providecommand \citenamefont [1]{#1}%
\providecommand \href@noop [0]{\@secondoftwo}%
\providecommand \href [0]{\begingroup \@sanitize@url \@href}%
\providecommand \@href[1]{\@@startlink{#1}\@@href}%
\providecommand \@@href[1]{\endgroup#1\@@endlink}%
\providecommand \@sanitize@url [0]{\catcode `\\12\catcode `\$12\catcode
  `\&12\catcode `\#12\catcode `\^12\catcode `\_12\catcode `\%12\relax}%
\providecommand \@@startlink[1]{}%
\providecommand \@@endlink[0]{}%
\providecommand \url  [0]{\begingroup\@sanitize@url \@url }%
\providecommand \@url [1]{\endgroup\@href {#1}{\urlprefix }}%
\providecommand \urlprefix  [0]{URL }%
\providecommand \Eprint [0]{\href }%
\providecommand \doibase [0]{http://dx.doi.org/}%
\providecommand \selectlanguage [0]{\@gobble}%
\providecommand \bibinfo  [0]{\@secondoftwo}%
\providecommand \bibfield  [0]{\@secondoftwo}%
\providecommand \translation [1]{[#1]}%
\providecommand \BibitemOpen [0]{}%
\providecommand \bibitemStop [0]{}%
\providecommand \bibitemNoStop [0]{.\EOS\space}%
\providecommand \EOS [0]{\spacefactor3000\relax}%
\providecommand \BibitemShut  [1]{\csname bibitem#1\endcsname}%
\let\auto@bib@innerbib\@empty


\bibitem [{\citenamefont {Castellanos-Gomez}\ \emph {et~al.}(2014)\citenamefont
  {Castellanos-Gomez}, \citenamefont {Buscema}, \citenamefont {Molenaar},
  \citenamefont {Singh}, \citenamefont {Janssen}, \citenamefont {Van
  Der~Zant},\ and\ \citenamefont
  {Steele}}]{SI_Castellanos-Gomez2014DeterministicStamping}%
  \BibitemOpen
  \bibfield  {author} {\bibinfo {author} {\bibfnamefont {A.}~\bibnamefont
  {Castellanos-Gomez}}, \bibinfo {author} {\bibfnamefont {M.}~\bibnamefont
  {Buscema}}, \bibinfo {author} {\bibfnamefont {R.}~\bibnamefont {Molenaar}},
  \bibinfo {author} {\bibfnamefont {V.}~\bibnamefont {Singh}}, \bibinfo
  {author} {\bibfnamefont {L.}~\bibnamefont {Janssen}}, \bibinfo {author}
  {\bibfnamefont {H.~S.}\ \bibnamefont {Van Der~Zant}}, \ and\ \bibinfo
  {author} {\bibfnamefont {G.~A.}\ \bibnamefont {Steele}},\ }\href {\doibase
  10.1088/2053-1583/1/1/011002} {\bibfield  {journal} {\bibinfo  {journal} {2D
  Materials}\ }\textbf {\bibinfo {volume} {1}},\ \bibinfo {pages} {011002}
  (\bibinfo {year} {2014})}\BibitemShut {NoStop}%
\bibitem [{\citenamefont {Sun}\ \emph {et~al.}(2020)\citenamefont {Sun},
  \citenamefont {Yang}, \citenamefont {Kuang}, \citenamefont {Stebunov},
  \citenamefont {Xiong}, \citenamefont {Yu}, \citenamefont {Nair},
  \citenamefont {Katsnelson}, \citenamefont {Yuan}, \citenamefont {Grigorieva},
  \citenamefont {Lozada-Hidalgo}, \citenamefont {Wang},\ and\ \citenamefont
  {Geim}}]{SI_Sun2020LimitsGraphene}%
  \BibitemOpen
  \bibfield  {author} {\bibinfo {author} {\bibfnamefont {P.~Z.}\ \bibnamefont
  {Sun}}, \bibinfo {author} {\bibfnamefont {Q.}~\bibnamefont {Yang}}, \bibinfo
  {author} {\bibfnamefont {W.~J.}\ \bibnamefont {Kuang}}, \bibinfo {author}
  {\bibfnamefont {Y.~V.}\ \bibnamefont {Stebunov}}, \bibinfo {author}
  {\bibfnamefont {W.~Q.}\ \bibnamefont {Xiong}}, \bibinfo {author}
  {\bibfnamefont {J.}~\bibnamefont {Yu}}, \bibinfo {author} {\bibfnamefont
  {R.~R.}\ \bibnamefont {Nair}}, \bibinfo {author} {\bibfnamefont {M.~I.}\
  \bibnamefont {Katsnelson}}, \bibinfo {author} {\bibfnamefont {S.~J.}\
  \bibnamefont {Yuan}}, \bibinfo {author} {\bibfnamefont {I.~V.}\ \bibnamefont
  {Grigorieva}}, \bibinfo {author} {\bibfnamefont {M.}~\bibnamefont
  {Lozada-Hidalgo}}, \bibinfo {author} {\bibfnamefont {F.~C.}\ \bibnamefont
  {Wang}}, \ and\ \bibinfo {author} {\bibfnamefont {A.~K.}\ \bibnamefont
  {Geim}},\ }\href {\doibase 10.1038/s41586-020-2070-x} {\bibfield  {journal}
  {\bibinfo  {journal} {Nature}\ }\textbf {\bibinfo {volume} {579}},\ \bibinfo
  {pages} {229} (\bibinfo {year} {2020})}\BibitemShut {NoStop}%
\bibitem [{\citenamefont {Kim}\ \emph {et~al.}(2018{\natexlab{b}})\citenamefont
  {Kim}, \citenamefont {Yu},\ and\ \citenamefont {van~der
  Zande}}]{SI_Kim2018Nano-electromechanicalBimorphs}%
  \BibitemOpen
  \bibfield  {author} {\bibinfo {author} {\bibfnamefont {S.}~\bibnamefont
  {Kim}}, \bibinfo {author} {\bibfnamefont {J.}~\bibnamefont {Yu}}, \ and\
  \bibinfo {author} {\bibfnamefont {A.~M.}\ \bibnamefont {van~der Zande}},\
  }\href {\doibase 10.1021/acs.nanolett.8b01926} {\bibfield  {journal}
  {\bibinfo  {journal} {Nano Letters}\ }\textbf {\bibinfo {volume} {18}},\
  \bibinfo {pages} {6686} (\bibinfo {year} {2018}{\natexlab{b}})}\BibitemShut
  {NoStop}%
\bibitem [{\citenamefont {Wang}\ \emph
  {et~al.}(2019{\natexlab{b}})\citenamefont {Wang}, \citenamefont {Dai},
  \citenamefont {Xiao}, \citenamefont {Feng}, \citenamefont {Weng},
  \citenamefont {Liu}, \citenamefont {Xu}, \citenamefont {Huang},\ and\
  \citenamefont {Zhang}}]{SI_Wang2019BendingMaterials}%
  \BibitemOpen
  \bibfield  {author} {\bibinfo {author} {\bibfnamefont {G.}~\bibnamefont
  {Wang}}, \bibinfo {author} {\bibfnamefont {Z.}~\bibnamefont {Dai}}, \bibinfo
  {author} {\bibfnamefont {J.}~\bibnamefont {Xiao}}, \bibinfo {author}
  {\bibfnamefont {S.}~\bibnamefont {Feng}}, \bibinfo {author} {\bibfnamefont
  {C.}~\bibnamefont {Weng}}, \bibinfo {author} {\bibfnamefont {L.}~\bibnamefont
  {Liu}}, \bibinfo {author} {\bibfnamefont {Z.}~\bibnamefont {Xu}}, \bibinfo
  {author} {\bibfnamefont {R.}~\bibnamefont {Huang}}, \ and\ \bibinfo {author}
  {\bibfnamefont {Z.}~\bibnamefont {Zhang}},\ }\href {\doibase
  10.1103/PhysRevLett.123.116101} {\bibfield  {journal} {\bibinfo  {journal}
  {Phys. Rev. Lett.}\ }\textbf {\bibinfo {volume} {123}},\ \bibinfo {pages}
  {116101} (\bibinfo {year} {2019}{\natexlab{b}})}\BibitemShut {NoStop}%
\bibitem [{\citenamefont {Falin}\ \emph {et~al.}(2017)\citenamefont {Falin},
  \citenamefont {Cai}, \citenamefont {Santos}, \citenamefont {Scullion},
  \citenamefont {Qian}, \citenamefont {Zhang}, \citenamefont {Yang},
  \citenamefont {Huang}, \citenamefont {Watanabe}, \citenamefont {Taniguchi}
  \emph {et~al.}}]{SI_falin2017mechanical}%
  \BibitemOpen
  \bibfield  {author} {\bibinfo {author} {\bibfnamefont {A.}~\bibnamefont
  {Falin}}, \bibinfo {author} {\bibfnamefont {Q.}~\bibnamefont {Cai}}, \bibinfo
  {author} {\bibfnamefont {E.~J.}\ \bibnamefont {Santos}}, \bibinfo {author}
  {\bibfnamefont {D.}~\bibnamefont {Scullion}}, \bibinfo {author}
  {\bibfnamefont {D.}~\bibnamefont {Qian}}, \bibinfo {author} {\bibfnamefont
  {R.}~\bibnamefont {Zhang}}, \bibinfo {author} {\bibfnamefont
  {Z.}~\bibnamefont {Yang}}, \bibinfo {author} {\bibfnamefont {S.}~\bibnamefont
  {Huang}}, \bibinfo {author} {\bibfnamefont {K.}~\bibnamefont {Watanabe}},
  \bibinfo {author} {\bibfnamefont {T.}~\bibnamefont {Taniguchi}},  \emph
  {et~al.},\ }\href {\doibase 10.1038/ncomms15815} {\bibfield  {journal}
  {\bibinfo  {journal} {Nature communications}\ }\textbf {\bibinfo {volume}
  {8}},\ \bibinfo {pages} {1} (\bibinfo {year} {2017})}\BibitemShut {NoStop}%
\bibitem [{\citenamefont {Dai}\ \emph {et~al.}(2020{\natexlab{b}})\citenamefont
  {Dai}, \citenamefont {Lu}, \citenamefont {Liechti},\ and\ \citenamefont
  {Huang}}]{SI_dai2020mechanicsopportunities}%
  \BibitemOpen
  \bibfield  {author} {\bibinfo {author} {\bibfnamefont {Z.}~\bibnamefont
  {Dai}}, \bibinfo {author} {\bibfnamefont {N.}~\bibnamefont {Lu}}, \bibinfo
  {author} {\bibfnamefont {K.~M.}\ \bibnamefont {Liechti}}, \ and\ \bibinfo
  {author} {\bibfnamefont {R.}~\bibnamefont {Huang}},\ }\href {\doibase
  10.1016/j.cossms.2020.100837} {\bibfield  {journal} {\bibinfo  {journal}
  {Current Opinion in Solid State and Materials Science}\ }\textbf {\bibinfo
  {volume} {24}},\ \bibinfo {pages} {100837} (\bibinfo {year}
  {2020}{\natexlab{b}})}\BibitemShut {NoStop}%
\end{thebibliography}
\end{document}